\documentclass[useAMS,usenatbib]{mn2e}

\usepackage{times}
\usepackage{graphicx}
\usepackage{multirow}
\usepackage{amsmath}

\def\pcm3{{\rm\thinspace cm^{-3}}}

\def\contcaption{\@conttrue\SFB@caption\@captype}

\newcommand{\prob}{\mbox {$\rm \surd$}}
\newcommand{\nae}{\mbox {$\rm \times$}}

\def\n_h{{\rm n_{H}}}

\def\NH1{{$N_{\rm HI}~$}}

\def\ga{{\rm\thinspace gauss}}

\def\approxlt{\mathrel{\hbox{\rlap{\lower .5ex \hbox {$\sim$}}
        \raise .15 ex \hbox{$<$}}}}
\def\approxgt{\mathrel{\hbox{\rlap{\lower .5ex \hbox {$\sim$}}
        \raise .15 ex \hbox{$>$}}}}

\def\la{\mathrel{\hbox{\rlap{\hbox{\lower4pt\hbox{$\sim$}}}\hbox{$<$}}}}
\def\ga{\mathrel{\hbox{\rlap{\hbox{\lower4pt\hbox{$\sim$}}}\hbox{$>$}}}}
\newbox\grsign \setbox\grsign=\hbox{$>$} \newdimen\grdimen
\grdimen=\ht\grsign
\newbox\simlessbox \newbox\simgreatbox \newbox\simpropbox
\setbox\simgreatbox=\hbox{\raise.5ex\hbox{$>$}\llap
     {\lower.5ex\hbox{$\sim$}}}\ht1=\grdimen\dp1=0pt
\setbox\simlessbox=\hbox{\raise.5ex\hbox{$<$}\llap
     {\lower.5ex\hbox{$\sim$}}}\ht2=\grdimen\dp2=0pt
\setbox\simpropbox=\hbox{\raise.5ex\hbox{$\propto$}\llap
     {\lower.5ex\hbox{$\sim$}}}\ht2=\grdimen\dp2=0pt
\def\simgreat{\mathrel{\copy\simgreatbox}}
\def\simless{\mathrel{\copy\simlessbox}}

\title[Non-magnetic white dwarf binaries in SDSS DR7]{Component masses of young, wide, non-magnetic white dwarf binaries in the SDSS DR7}

\author[Baxter et al.]{R. B. Baxter$^{1}$\thanks{E-mail: richardbrucebaxter@gmail.com}, P. D. Dobbie$^{2}$, Q. A. Parker$^{1,3,4}$, S.L. Casewell$^{5}$,  N. Lodieu$^{6,7}$, M.R. Burleigh$^{5}$, \newauthor  K.A.Lawrie$^{5}$, B. K\"ulebi$^{8}$, D. Koester$^{9}$, B. R. Holland$^{2}$  \\
$^{1}$Dept. of Physics \& Astronomy, Macquarie University, NSW, 2109, Australia\\
$^{2}$School of Physical Sciences, University of Tasmania, Hobart, TAS, 7001, Australia\\
$^{3}$Macquarie Research Center for Astronomy, Astrophysics \& Astrophotonics, Macquarie University, NSW, 2109, Australia\\
$^{4}$Australian Astronomical Observatory, PO Box 915, North Ryde, NSW, 1670, Australia \\
$^{5}$Dept. of Physics \& Astronomy, University of Leicester, Leicester, LE1 7RH, UK\\
$^{6}$Instituto de Astrofisica de Canarias, Via Lactea s/n, E-38200 La Laguna, Tenerife, Spain\\
$^{7}$Departamento de Astrof\'isica, Universidad de La Laguna (ULL), E-38206 La Laguna, Tenerife, Spain \\
$^{8}$Institut de Ci\`{e}ncies de l$^{\prime}$Espai (CSIC-IEEC), Facultat de Ci\`{e}ncies, Campus UAB, Torre C5-parell, 2$^{\rm a}$  planta, 08193 Bellaterra, Spain \\
$^{9}$Institut f\"ur Theoretische Physik und Astrophysik, Christian-Albrechts-Universit\"at, Kiel, Germany \\
}

\begin{document}

\date{Accepted . Received ; in original form }

\pagerange{\pageref{firstpage}--\pageref{lastpage}} \pubyear{2009}

\maketitle

\label{firstpage}

\begin{abstract}

 We present a spectroscopic component analysis of 18 candidate young, wide, non-magnetic, double-degenerate binaries identified from 
a search of the Sloan Digital Sky Survey Data Release 7 (DR7). All but two pairings are likely to be physical systems. We show 
SDSS\,J084952.47+471247.7 + SDSS\,J084952.87+471249.4 to be a wide DA + DB binary, only the second identified to date. Combining our 
measurements for the components of 16 new binaries with results for three similar, previously known systems within the DR7, we have 
constructed a mass distribution for the largest sample to date (38) of white dwarfs in young, wide, non-magnetic, double-degenerate 
pairings. This is broadly similar in form to that of the isolated field population with a substantial peak around $M$$\sim$0.6$M_{\odot}$.
We identify an excess of ultra-massive white dwarfs and attribute this to the primordial separation distribution of their progenitor 
systems peaking at relatively larger values and the greater expansion of their binary orbits during the final stages of stellar evolution. 
We exploit this mass distribution to probe the origins of unusual types of degenerates, confirming a mild preference for the progenitor 
systems of high-field-magnetic white dwarfs, at least within these binaries, to be associated with early-type stars. Additionally, we 
consider the 19 systems in the context of the stellar initial mass-final mass relation. None appear to be strongly discordant with current
understanding of this relationship.

\end{abstract}


\section{Introduction}

\begin{figure}
\includegraphics[angle=0,width=\linewidth]{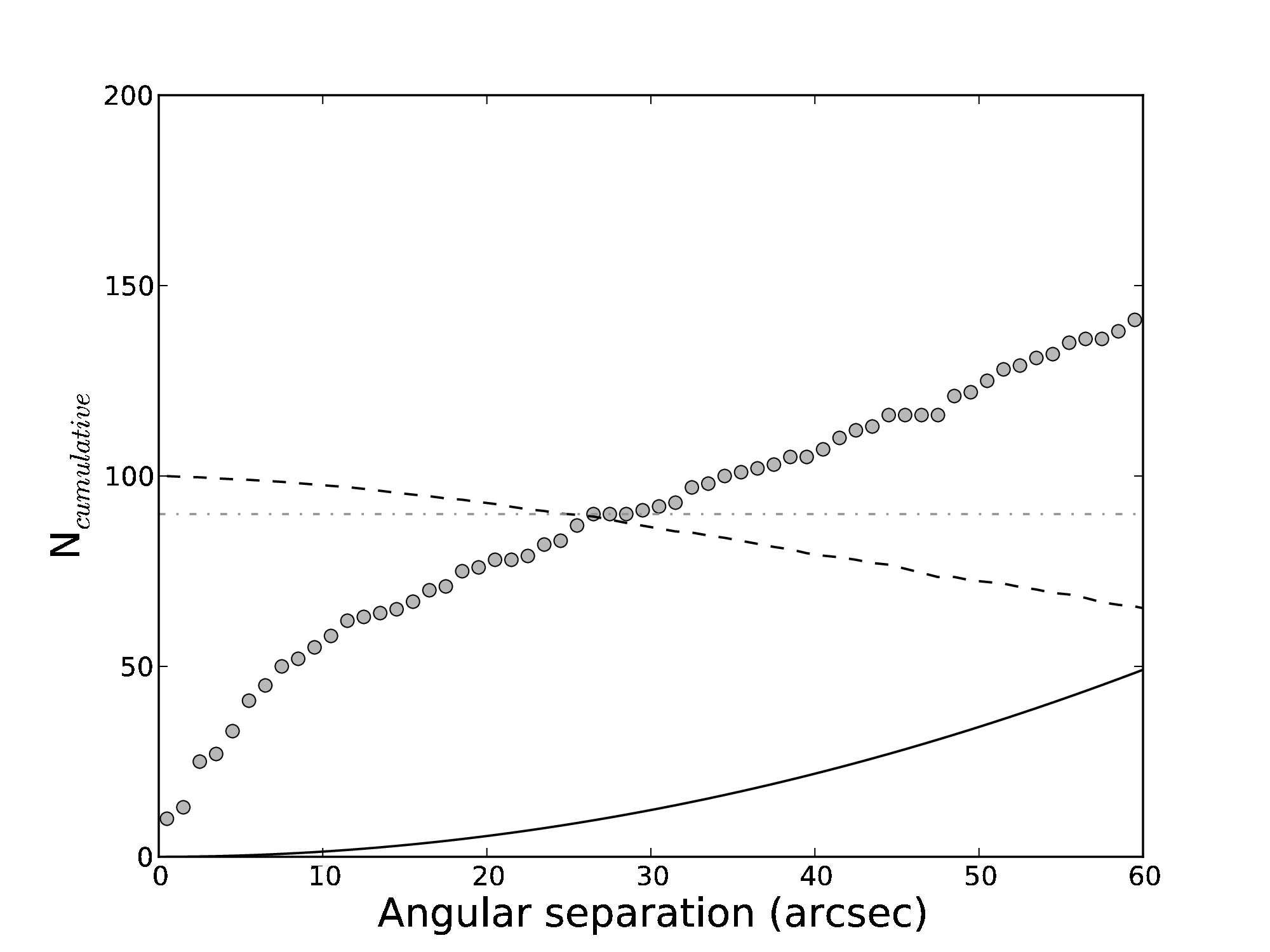}
\caption{A plot of the cumulative number of observed pairings that meet our photometric selection criteria (filled grey circles) and are expected 
for a random on-sky distribution of objects (black line) as a function of angular separation. A crude estimate of the proportion of physical systems 
as a function of angular separation is also shown (dashed line). For angular separations of 30 arcsec or less, roughly 90\% of candidates are likely 
to be physical binaries.}
\label{cumulate}
\end{figure}

A substantial proportion of stars reside in binary or multiple stellar systems \citep[e.g.][]{duquennoy91, fischer92, kouwenhoven05, kouwenhoven07b}. Empirical 
determinations of the stellar binary fraction as a function of primary mass and of the binary mass ratio and orbital period distributions inform theories of the
 star formation process \citep[e.g.][]{zinnecker84, pinfield03, parker13}. Moroever, studies of close systems, with orbital periods of a 
few days or less, can yield important dynamical determinations of masses and radii which lend themselves to arguably the most stringent examinations 
of models of stellar structure \citep[e.g.][]{huang56, maxted04, clausen08}. Wide, spatially resolved, binary systems where the components are separated by 
 100--10000AU and have generally evolved essentially as single stars \citep[e.g.][]{andrews12}, are also of significant interest, since they are, in effect, miniature versions of open 
clusters, the traditional but often rather distant testbeds for refining our theories of stellar evolution 
\citep[e.g.][]{barbaro84,nordstrom96, casewell09, kalirai10, casewell12}.

By considering observational constraints on the stellar binary fraction across a broad range of primary masses, from the late-F/G stars \citep[57\%][]{duquennoy91} 
to the numerically dominant low mass M dwarfs \citep[26\%, ][]{delfosse04}, \cite{lada06b} has highlighted that the majority of stars reside in single stellar 
systems. However, since only those stars of the Galactic disk with $M$$\simgreat$1$M$$_{\odot}$ have had sufficient time to evolve beyond the main-sequence, around 
half or more of the members of the field white dwarf population presumably must have once been part of multiple systems. \cite{miszalski09} determine that around least 
12--21\% of planetary nebulae have close binary central stars, while  \cite{holberg08b, holberg13}  have concluded that at least 25-30\%  of the white dwarfs within 20pc 
of the Sun are presently part of binary systems, with around 6\% being double-degenerates. The systems at the short end of the double-degenerate period distribution are of substantial 
astrophysical relevance since a subset may ultimately evolve to Type Ia supernovae \citep[e.g.][]{yoon07}. Widely separated double-degenerates are also of interest 
for setting limits on the age of the Galactic disk through the white dwarf luminosity function \citep{oswalt96} and for investigating the late-stages of stellar 
evolution, in particular the heavy mass-loss experienced on the asymptotic giant branch, as manifest through the form of the stellar initial-mass final-mass relation 
\citep[e.g.][]{finley97}. Additionally, when a wide double-degenerate harbours an unusual or peculiar white dwarf (e.g. a high field magnetic white dwarf), 
measurements of the parameters of the normal component can be used to investigate its fundamental parameters and origins, either directly or potentially statistically
\citep[e.g.][]{girven10,dobbie12a,dobbie13a}. 
 
Here we begin to build the foundations for a statistical approach by presenting the mass distribution for what is by far the largest spectroscopically observed sample 
(38) of non-magnetic white dwarfs residing in young, wide, double-degenerate systems, to date. In subsequent sections we describe our photometric identification of 53 
candidate young, wide, binary systems from a search of the Sloan Digital Sky Survey (SDSS) data release 7 \citep[DR7, ][]{abazajian09} and discuss our new 
spectroscopic follow-up and analysis of the components of 18 of these systems and detail our assessment of their physical reality. We assemble a mass distribution 
for the components of the systems we find to have a strong likelihood of being binaries and those of three previously identified wide double-degenerates within the
DR7 footprint. We compare this to that of isolated field white dwarfs and discuss the similarities and the differences. We demonstate how this mass distribution can 
be used to probe the origins of unsual degenerates, in this case high-field-magentic white dwarfs (HFMWDs). Finally, we explore the 19 non-magnetic binary systems 
within the context of our current understanding of the form of the stellar initial mass-final mass relation.

\begin{table*}
\setlength{\tabcolsep}{4pt}.
\scriptsize
 \centering
 \begin{minipage}{180mm}
  \caption{Survey designation, SDSS designation, photometric data and observed angular separations for the candidate double-degenerate systems (CDDS) we have identified. Candidates for which we have obtained new resolved spectroscopy (Spec follow-up = Y) and objects for which a spectroscopic analysis exists in the literature (Spec follow-up = L), are labelled. White dwarfs included in {\protect \cite{baxter11}} (B) and those with SDSS DR7 spectroscopy (italics) are also highlighted.}
  \label{allcands}  
  \begin{tabular}{lcccccccccccc}
  \hline
   CDDS ID & SDSS  &  Spec & $u$  & $g$ & $r$ & $i$ & SDSS & $u$ & $g$ & $r$ & $i$ & Sep. \\
  (Desig.)  & (Desig.) & follow-up & (/mag.) & (/mag.) & (/mag.) & (/mag.) & (Desig.) & (/mag.) & (/mag.) & (/mag.) & (/mag.) & (/arcsec) \\


\hline
CDDS1 &   J000142.84+251506.1 & 	&17.81(0.02) & 17.79(0.02) & 18.16(0.02) & 18.45(0.02) &	{\it J000142.79+251504.0} &  	19.16(0.19) & 18.70(0.16) & 19.01(0.16) & 19.32(0.17) &	   2.16  \\ 
CDDS2 &   J002925.28+001559.7 & 	&20.02(0.05) & 19.59(0.02) & 19.59(0.02) & 19.68(0.02) &	{\it J002925.62+001552.7} &  	18.91(0.03) & 18.48(0.01) & 18.53(0.02) & 18.65(0.02) &	   8.64  \\ 
CDDS3$^{B}$ &   J005212.26+135302.0 & Y	&17.79(0.02) & 17.71(0.03) & 17.98(0.02) & 18.24(0.02) &	J005212.73+135301.1 &  	19.35(0.03) & 18.89(0.03) & 18.92(0.02) & 19.05(0.02) &	   6.78  \\ 
CDDS4 &   {\it J011714.48+244021.5} & 	&19.94(0.03) & 19.63(0.02) & 19.77(0.02) & 20.06(0.03) &	J011714.12+244020.3 &  	20.29(0.04) & 19.83(0.02) & 19.96(0.02) & 20.16(0.03) &	   5.05  \\ 
CDDS5 &   J012726.89+391503.3 & 	&19.16(0.03) & 18.70(0.02) & 18.83(0.02) & 19.01(0.02) &	J012725.51+391459.2 &  	20.35(0.06) & 19.99(0.02) & 19.99(0.02) & 20.07(0.03) &	  16.55  \\ 
CDDS6$^{B}$ &   J021131.51+171430.4 & Y	&17.36(0.12) & 17.26(0.09) & 17.65(0.07) & 17.87(0.08) &	J021131.52+171428.3 &  	16.60(0.02) & 16.71(0.02) & 17.08(0.01) & 17.40(0.01) &	   2.04  \\ 
CDDS7 &   {\it J022733.09+005200.3} & 	&20.03(0.05) & 19.62(0.02) & 19.69(0.02) & 19.80(0.03) &	{\it J022733.15+005153.6} &  	20.30(0.06) & 19.86(0.02) & 19.91(0.02) & 19.99(0.03) &	   6.72  \\ 
CDDS8$^{B}$ &   {\it J033236.86-004936.9} & Y	&15.32(0.01) & 15.64(0.02) & 16.09(0.02) & 16.41(0.02) &	{\it J033236.60-004918.4} &  	18.64(0.03) & 18.20(0.02) & 18.30(0.02) & 18.45(0.02) &	  18.91  \\ 
CDDS9 &   J054519.81+302754.0 & Y	&20.19(0.05) & 19.82(0.02) & 19.96(0.02) & 20.10(0.03) &	J054518.98+302749.3 &  	20.05(0.05) & 19.64(0.02) & 19.80(0.02) & 19.97(0.03) &	  11.72  \\ 
CDDS10 &  {\it J072147.38+322824.1} & 	&18.18(0.02) & 18.07(0.01) & 18.17(0.01) & 18.28(0.01) &	J072147.20+322822.4 &  	18.76(0.08) & 18.32(0.06) & 18.34(0.06) & 18.44(0.06) &	   2.77  \\ 
CDDS11$^{D2}$ &  {\it J074853.07+302543.5} &  Y	&17.41(0.07) & 17.59(0.05) & 17.88(0.05) & 18.49(0.12) &	J074852.95+302543.4 &  	17.57(0.05) & 17.59(0.04) & 17.96(0.04) & 18.24(0.03) &	   1.50  \\ 
CDDS12 &  J075410.53+123947.3 & 	&19.19(0.09) & 18.78(0.06) & 18.99(0.08) & 19.22(0.07) &	J075410.58+123945.5 &  	19.22(0.04) & 18.86(0.05) & 18.99(0.04) & 19.25(0.04) &	   2.00  \\ 
CDDS13 &  {\it J080212.54+242043.6} & 	&19.87(0.04) & 19.54(0.02) & 19.78(0.02) & 20.01(0.03) &	J080213.44+242020.9 &  	20.23(0.04) & 19.84(0.02) & 20.00(0.02) & 20.19(0.03) &	  25.85  \\ 
CDDS14 &  J080644.09+444503.2 & Y	&18.54(0.02) & 18.14(0.02) & 18.32(0.01) & 18.54(0.02) &	J080643.64+444501.4 &  	19.18(0.02) & 18.74(0.02) & 18.82(0.01) & 18.94(0.02) &	   5.09  \\ 
CDDS15 &  J084952.87+471249.4 & Y	&16.64(0.02) & 16.77(0.02) & 17.08(0.02) & 17.35(0.02) &	J084952.47+471247.7 &  	18.14(0.14) & 17.77(0.08) & 17.79(0.06) & 18.04(0.07) &	   4.37  \\ 
CDDS16 &  J085915.02+330644.6 & Y	&18.27(0.02) & 18.01(0.02) & 18.34(0.02) & 18.59(0.02) &	J085915.50+330637.6 &  	19.07(0.03) & 18.70(0.02) & 18.87(0.02) & 19.04(0.02) &	   9.29  \\ 
CDDS17 &  J085917.36+425031.6 & Y	&19.37(0.03) & 18.94(0.05) & 19.01(0.02) & 19.08(0.02) &	J085917.23+425027.4 &  	18.83(0.02) & 18.38(0.04) & 18.53(0.02) & 18.70(0.02) &	   4.39  \\ 
CDDS18$^{*}$ &  J092513.18+160145.4 & L &17.07(0.09) & 17.12(0.08) & 17.52(0.06) & 17.83(0.06) &	J092513.48+160144.1 &  	16.07(0.02) & 16.14(0.02) & 16.55(0.01) & 16.88(0.02) &	   4.51  \\ 
CDDS19$^{D1}$ &  J092647.00+132138.4 & Y	&18.74(0.03) & 18.40(0.03) & 18.46(0.05) & 18.60(0.04) &	J092646.88+132134.5 &  	18.46(0.02) & 18.34(0.02) & 18.39(0.02) & 18.50(0.02) &	   4.35  \\ 
CDDS20 &  J095458.73+390104.6 & 	&20.31(0.05) & 19.86(0.02) & 19.95(0.03) & 19.95(0.03) &	{\it J095459.97+390052.4} &  	17.96(0.02) & 17.69(0.02) & 18.02(0.02) & 18.29(0.02) &	  18.87  \\ 
CDDS21 &  J100245.86+360653.3 & 	&19.42(0.03) & 19.04(0.02) & 19.09(0.02) & 19.16(0.03) &	{\it J100244.88+360629.6} &  	19.32(0.03) & 18.92(0.02) & 19.01(0.02) & 19.09(0.03) &	  26.53  \\ 
CDDS22$^{D3}$ &  J105306.13+025052.5 & Y	&19.57(0.04) & 19.14(0.02) & 19.28(0.02) & 19.51(0.03) &	{\it J105306.82+025027.9} &  	19.37(0.03) & 18.98(0.02) & 19.18(0.02) & 19.37(0.03) &	  26.60  \\ 
CDDS23 &  J113928.52-001420.9 & 	&19.84(0.04) & 19.42(0.02) & 19.52(0.02) & 19.71(0.03) &	J113928.47-001418.0 &  	20.13(0.07) & 19.80(0.06) & 19.85(0.06) & 19.93(0.08) &	   2.95  \\ 
CDDS24 &  J115030.12+253210.1 & 	&20.43(0.05) & 19.95(0.02) & 19.97(0.02) & 20.05(0.04) &	J115030.48+253206.0 &  	19.30(0.03) & 18.86(0.02) & 19.09(0.02) & 19.29(0.02) &	   6.38  \\ 
CDDS25 &  {\it J115305.54+005646.1} & 	&18.42(0.02) & 18.89(0.02) & 19.38(0.02) & 19.62(0.03) &	J115305.47+005645.8 &  	18.50(0.02) & 18.91(0.02) & 19.34(0.02) & 19.78(0.03) &	   1.22  \\ 
CDDS26$^{B}$ &  {\it J115937.81+134413.9} & Y	&18.45(0.03) & 18.07(0.02) & 18.12(0.02) & 18.16(0.02) &	J115937.82+134408.7 &  	18.42(0.03) & 18.28(0.02) & 18.52(0.02) & 18.75(0.03) &	   5.18  \\ 
CDDS27$^{D3}$ &  J122739.16+661224.4 & Y	&17.72(0.02) & 17.86(0.02) & 18.13(0.02) & 18.44(0.02) &	J122741.05+661224.3 &  	18.23(0.02) & 17.99(0.02) & 18.21(0.02) & 18.46(0.02) &	  11.43  \\ 
CDDS28 &  {\it J131012.28+444728.3} & 	&17.88(0.01) & 17.84(0.02) & 18.02(0.01) & 18.23(0.02) &	J131013.38+444717.8 &  	17.95(0.01) & 17.59(0.02) & 17.85(0.01) & 18.11(0.02) &	  15.71  \\ 
CDDS29$^{B}$ &  J131332.14+203039.6 & Y	&18.13(0.02) & 17.80(0.02) & 17.98(0.01) & 18.19(0.02) &	J131332.56+203039.3 &  	17.86(0.02) & 17.48(0.02) & 17.69(0.01) & 17.91(0.02) &	   5.93  \\ 
CDDS30 &  J131421.70+305051.4 & Y	&18.59(0.10) & 18.20(0.08) & 18.22(0.09) & 18.31(0.09) &	{\it J131421.50+305050.5} &  	18.23(0.04) & 17.86(0.04) & 17.88(0.05) & 18.01(0.06) &	   2.76  \\ 
CDDS31$^{B}$ &  J132814.28+163151.5 & Y	&16.34(0.02) & 16.27(0.02) & 16.63(0.02) & 16.99(0.02) &	J132814.36+163150.9 &  	17.75(0.27) & 17.65(0.23) & 17.74(0.19) & 17.84(0.15) &	   1.32  \\ 
CDDS32 &  J135713.14-065913.7 & Y	&18.94(0.04) & 19.25(0.02) & 19.76(0.02) & 20.18(0.04) &	J135714.50-065856.9 &  	18.58(0.04) & 18.16(0.02) & 18.35(0.02) & 18.54(0.02) &	  26.29  \\ 
CDDS33$^{D1}$ &  J150746.48+521002.1 & Y	&17.14(0.02) & 16.91(0.03) & 17.29(0.01) & 17.55(0.02) &	J150746.80+520958.0 &  	17.98(0.02) & 17.76(0.03) & 18.06(0.01) & 18.33(0.02) &	   5.05  \\ 
CDDS34 &  J151508.30+143640.8 & Y	&18.38(0.02) & 18.00(0.02) & 18.20(0.01) & 18.47(0.02) &	J151507.90+143635.4 &  	19.76(0.03) & 19.63(0.02) & 19.88(0.02) & 20.19(0.03) &	   7.90  \\ 
CDDS35 &  J154641.48+615901.7 & 	&19.07(0.03) & 18.63(0.02) & 18.75(0.02) & 18.93(0.02) &	J154641.79+615854.3 &  	17.16(0.02) & 16.89(0.02) & 17.17(0.02) & 17.42(0.02) &	   7.64  \\ 
CDDS36 &  J155245.19+473129.5 & Y	&18.79(0.02) & 18.71(0.02) & 19.06(0.02) & 19.36(0.03) &	J155244.41+473124.0 &  	19.21(0.04) & 18.99(0.03) & 19.30(0.02) & 19.61(0.03) &	   9.65  \\ 
CDDS37 &  J162650.11+482827.9 & 	&19.72(0.03) & 19.62(0.02) & 19.94(0.03) & 20.20(0.04) &	J162652.12+482824.7 &  	19.14(0.02) & 18.98(0.01) & 19.30(0.02) & 19.59(0.02) &	  20.22  \\ 
CDDS38 &  J163647.81+092715.7 & 	&18.13(0.02) & 17.72(0.01) & 17.93(0.01) & 18.18(0.01) &	J163647.33+092708.4 &  	19.98(0.04) & 19.54(0.02) & 19.54(0.02) & 19.66(0.03) &	  10.12  \\ 
CDDS39 &  {\it J165737.90+620102.1} & 	&18.72(0.02) & 18.65(0.01) & 18.98(0.03) & 19.23(0.02) &	J165734.39+620055.9 &  	18.88(0.02) & 18.53(0.01) & 18.76(0.02) & 18.99(0.02) &	  25.47  \\ 
CDDS40$^{B}$ &  {\it J170355.91+330438.4} & Y	&19.16(0.02) & 18.81(0.01) & 18.86(0.01) & 18.97(0.02) &	J170356.77+330435.7 &  	18.48(0.02) & 18.16(0.01) & 18.27(0.01) & 18.42(0.02) &	  11.16  \\ 
CDDS41 &  J173249.57+563900.0 & 	&19.35(0.03) & 18.95(0.02) & 19.12(0.02) & 19.35(0.02) &	J173249.32+563858.8 &  	18.99(0.04) & 19.12(0.05) & 19.27(0.06) & 19.47(0.04) &	   2.36  \\ 
CDDS42 &  J175559.57+484359.9 & 	&19.04(0.03) & 19.21(0.02) & 19.39(0.02) & 19.43(0.02) &	J175558.35+484348.8 &  	18.02(0.02) & 17.68(0.01) & 17.90(0.01) & 18.17(0.02) &	  16.41  \\ 
CDDS43 &  J204318.96+005841.8 & 	&18.51(0.03) & 18.24(0.02) & 18.42(0.01) & 18.59(0.02) &	J204317.93+005830.5 &  	18.96(0.03) & 18.59(0.02) & 18.75(0.01) & 18.94(0.02) &	  19.13  \\ 
CDDS44 &  J211607.27+004503.1 & 	&18.60(0.02) & 18.67(0.01) & 18.89(0.01) & 19.11(0.02) &	J211607.20+004501.3 &  	19.43(0.10) & 18.96(0.07) & 19.05(0.09) & 19.28(0.06) &	   2.06  \\ 
CDDS45 &  J213648.79+064320.2 & 	&18.07(0.02) & 17.94(0.02) & 18.24(0.01) & 18.48(0.02) &	J213648.98+064318.2 &  	19.72(0.04) & 19.35(0.02) & 19.39(0.03) & 19.50(0.02) &	   3.44  \\ 
CDDS46 &  J214456.12+482352.9 & 	&19.19(0.03) & 18.74(0.01) & 18.83(0.02) & 19.02(0.02) &	J214457.39+482345.5 &  	19.81(0.05) & 19.49(0.02) & 19.49(0.02) & 19.64(0.03) &	  14.67  \\ 
CDDS47 &  J215309.89+461902.7 & 	&18.15(0.02) & 17.72(0.01) & 17.90(0.01) & 18.05(0.01) &	J215308.90+461839.1 &  	18.88(0.03) & 19.08(0.01) & 19.36(0.02) & 19.56(0.02) &	  25.68  \\ 
CDDS48$^{B}$ &  J222236.30-082808.0 & Y	&16.68(0.02) & 16.41(0.02) & 16.67(0.03) & 16.92(0.03) &	J222236.56-082806.0 &  	17.56(0.03) & 17.11(0.07) & 17.30(0.07) & 17.47(0.06) &	   4.29  \\ 
CDDS49$^{+}$ &  J222301.62+220131.3 & L	&15.66(0.01) & 15.60(0.01) & 15.91(0.01) & 16.22(0.01) &	J222301.72+220124.9 &  	16.37(0.01) & 16.01(0.03) & 16.20(0.03) & 16.46(0.01) &	   6.56  \\ 
CDDS50$^{B}$ &  J222427.07+231537.4 & Y	&17.53(0.02) & 17.15(0.02) & 17.36(0.02) & 17.47(0.02) &	J222426.91+231536.0 &  	18.22(0.08) & 17.77(0.07) & 17.92(0.07) & 17.94(0.06) &	   2.64  \\ 
CDDS51$^{\dagger}$ & J224231.14+125004.9 & L &16.48(0.01) & 16.23(0.02) & 16.50(0.01) & 16.74(0.02) & J224230.33+125002.3 &  16.83(0.01) & 16.50(0.02) & 16.75(0.01) & 16.97(0.02) &	  12.13  \\ 
CDDS52$^{D3}$ &  J225932.74+140444.2 & Y	&19.02(0.03) & 18.57(0.02) & 18.68(0.01) & 18.85(0.02) &	J225932.21+140439.2 &  	16.16(0.02) & 16.36(0.01) & 16.78(0.01) & 17.12(0.01) &	   9.14  \\ 
CDDS53 &  J233246.27+491712.0 &	        &18.76(0.02) & 18.64(0.01) & 18.91(0.01) & 19.16(0.02) &	{\it J233246.23+491709.1} &  	19.02(0.06) & 18.76(0.04) & 19.04(0.04) & 19.31(0.05) &	   2.96  \\

\hline
\end{tabular} 

$^{*}$ PG\,0922+162A+B \citep{finley97} \\
$^{+}$ HS\,2220+2146A+B \citep{koester09}\\
$^{\dagger}$ HS\,2240+1234 \citep{jordan98} \\
$^{B}$ Preliminary analysis presented in  \cite{baxter11}\\
$^{D1,D2,D3}$ DA + DAH pairings discussed in  \citep{dobbie12a,dobbie13a} and Dobbie et al. (in prep), respectively.
\normalsize
\end{minipage}
\end{table*}

\begin{table*}
\begin{minipage}{175mm}
\begin{center}
\label{slog1}
\caption{Summary of our spectroscopic observations, including telescope/instrument combination and exposure times, of the candidate young, wide, double-degenerates within the SDSS DR7 imaging (RA=0--12h). }

\begin{tabular}{lccccccccc}
\hline
\multicolumn{1}{c}{ID} & SpT & SDSS & Telescope/Instrument & Exposure & N$_{exp}$ \\ 

\hline
CDDS3-A  & DA & J005212.73+135301.1   & \multirow{2}{*}{WHT + ISIS}  & \multirow{2}{*}{2400s} & \multirow{2}{*}{5}\\
CDDS3-B  & DA & J005212.26+135302.0   & & &\\ \\

CDDS6-A  & DA & J021131.52+171428.3   & \multirow{2}{*}{GEM-N + GMOS} & \multirow{2}{*}{2000s} & \multirow{2}{*}{3}\\
CDDS6-B  & DA & J021131.51+171430.4   & \\ \\

CDDS8-A  & DA & J033236.86-004936.9   & \multirow{2}{*}{VLT + FORS} & \multirow{2}{*}{600s} & \multirow{2}{*}{2}\\
CDDS8-B  & DA & J033236.60-004918.4   & \\ \\

CDDS9-A  & DA & J054519.81+302754.0   & \multirow{2}{*}{GTC + OSIRIS} & \multirow{2}{*}{2400s} & \multirow{2}{*}{3}  \\
CDDS9-B  & DA & J054518.98+302749.3   & & &\\ \\

CDDS14-A & DA & J080644.09+444503.2   & \multirow{2}{*}{GTC + OSIRIS} & \multirow{2}{*}{600s} & \multirow{2}{*}{3}\\
CDDS14-B & DA & J080643.64+444501.4   & & &\\ \\

CDDS15-A & DB & J084952.87+471249.4   & \multirow{2}{*}{GTC + OSIRIS}& \multirow{2}{*}{240s} & \multirow{2}{*}{3}\\
CDDS15-B & DA & J084952.47+471247.7   & & &\\ \\

CDDS16-A & DA & J085915.50+330637.6  & \multirow{2}{*}{GTC + OSIRIS} & \multirow{2}{*}{600s} & \multirow{2}{*}{3} \\
CDDS16-B & DA & J085915.02+330644.6  & & &\\ \\

CDDS17-A & DA & J085917.36+425031.6   & \multirow{2}{*}{GTC + OSIRIS} & \multirow{2}{*}{900s} & \multirow{2}{*}{3}\\
CDDS17-B & DA & J085917.23+425027.4   & & &\\ \\

CDDS18-A$^{*}$ & DA & J092513.48+160144.1   & \multirow{2}{*}{\cite{koester09}} & &\\
CDDS18-B$^{*}$ & DA & J092513.18+160145.4   & & & \\ \\

CDDS26-A & DA & J115937.82+134408.7 
  & \multirow{2}{*}{VLT + FORS} & \multirow{2}{*}{600s} & \multirow{2}{*}{2} \\
CDDS26-B & DA & J115937.81+134413.9   & & &\\ \\

\hline
\end{tabular}
\label{slog1}
\end{center}
$^{*}$ PG\,0922+162A+B \citep{finley97} \\
\end{minipage}
\label{slog1}
\end{table*}

\addtocounter{table}{-1}

\begin{table*}
\begin{minipage}{175mm}
\begin{center}
\label{tab3}
\caption{Summary of our spectroscopic observations, including telescope/instrument combination and exposure times, of the candidate young, wide, double-degenerates 
within the SDSS DR7 imaging (RA=12--24h).}
\begin{tabular}{lccccccccc}
\hline
\multicolumn{1}{c}{ID} & SpT & SDSS & Telescope/Instrument & Exposure & N$_{exp}$ \\ 

\hline


CDDS29-A & DA & J131332.56+203039.3 & \multirow{2}{*}{VLT + FORS} & \multirow{2}{*}{300s} & \multirow{2}{*}{2} \\
CDDS29-B & DA & J131332.14+203039.6 & & & \\ \\


CDDS30-A & DA & J131421.70+305051.4   & \multirow{2}{*}{VLT + FORS} & \multirow{2}{*}{600s} & \multirow{2}{*}{2}\\
CDDS30-B & DA & J131421.50+305050.5   & & &\\ \\


CDDS31-A & DA & J132814.36+163150.9   & \multirow{2}{*}{VLT + FORS} & \multirow{2}{*}{300s} & \multirow{2}{*}{2}\\
CDDS31-B & DA & J132814.28+163151.5   & & &\\ \\

CDDS32-A & DA & J135714.50-065856.9  & \multirow{2}{*}{VLT + FORS} & \multirow{2}{*}{600s} & \multirow{2}{*}{1}\\
CDDS32-B & sdO & J135713.14-065913.7  & & &\\ \\


CDDS34-A & DA & J151508.30+143640.8   & \multirow{2}{*}{GTC + OSIRIS} & \multirow{2}{*}{1800s} & \multirow{2}{*}{3}\\
CDDS34-B & DA & J151507.90+143635.4   & & &\\ \\


CDDS36-A & DA & J155245.19+473129.5   & \multirow{2}{*}{GTC + OSIRIS} & \multirow{2}{*}{900s} & \multirow{2}{*}{3}\\
CDDS36-B & DA & J155244.41+473124.0   &  & & \\ \\


CDDS40-A & DA & J170356.77+330435.7   & \multirow{2}{*}{WHT + ISIS} & \multirow{2}{*}{1800s} & \multirow{2}{*}{7} \\
CDDS40-B & DA & J170355.91+330438.4   & & &\\ \\


CDDS48-A & DA & J222236.56-082806.0   & \multirow{2}{*}{GEM-S + GMOS} & \multirow{2}{*}{1800s} & \multirow{2}{*}{3}\\
CDDS48-B & DA & J222236.30-082808.0   & & &\\ \\

CDDS49-A$^{+}$ & DA & J222301.72+220124.9   & \multirow{2}{*}{\cite{koester09}} & &  \\
CDDS49-B$^{+}$ & DA & J222301.62+220131.3   & & &\\ \\


CDDS50-A & DA & J222427.07+231537.4   & \multirow{2}{*}{WHT + ISIS} & \multirow{2}{*}{1200s} & \multirow{2}{*}{2}\\
CDDS50-B & DA & J222426.91+231536.0   & \\ \\

CDDS51-A$^{\dagger}$ & DA & J224231.14+125004.9  & \multirow{2}{*}{\cite{koester09}} & & \\
CDDS51-B$^{\dagger}$ & DA & J224230.33+125002.3  & & & \\ \\

\hline
\end{tabular}
\label{tab3}
\end{center}
$^{+}$ HS\,2220+2146A+B \citep{koester09}\\
$^{\dagger}$ HS\,2240+1234 \citep{jordan98}
\end{minipage}
\label{tab3}
\end{table*}

\begin{figure*}
\includegraphics[angle=0,width=12.5cm]{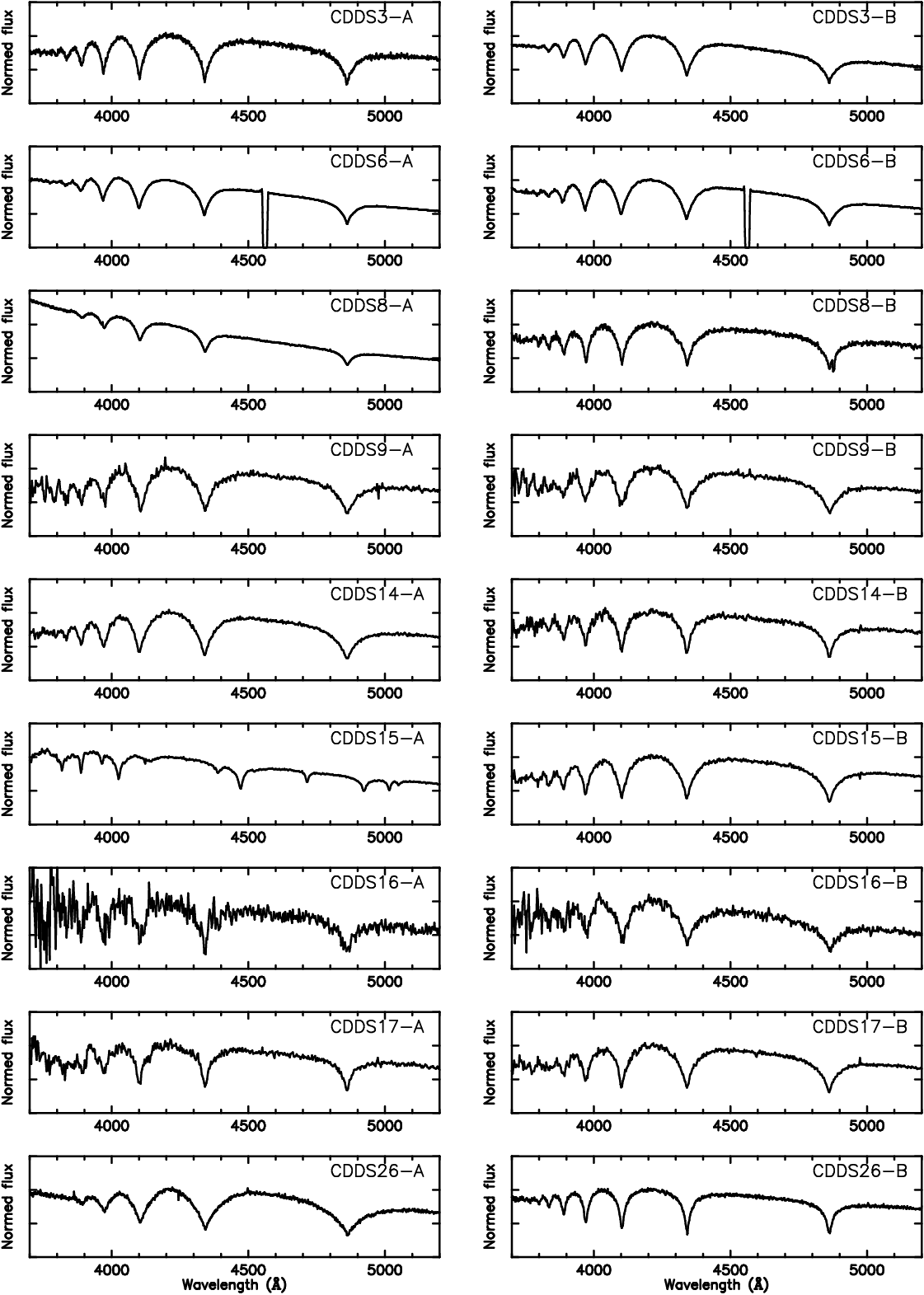}
\caption{Low resolution optical spectroscopy for the components of candidate binary systems in the range RA=0--12hr.  These data have been normalised by dividing by the median flux in the interval $\lambda$=4180 -- 4220\AA.}
\label{specs1}
\end{figure*}

\addtocounter{figure}{-1}

\begin{figure*}
\includegraphics[angle=0,width=12.5cm]{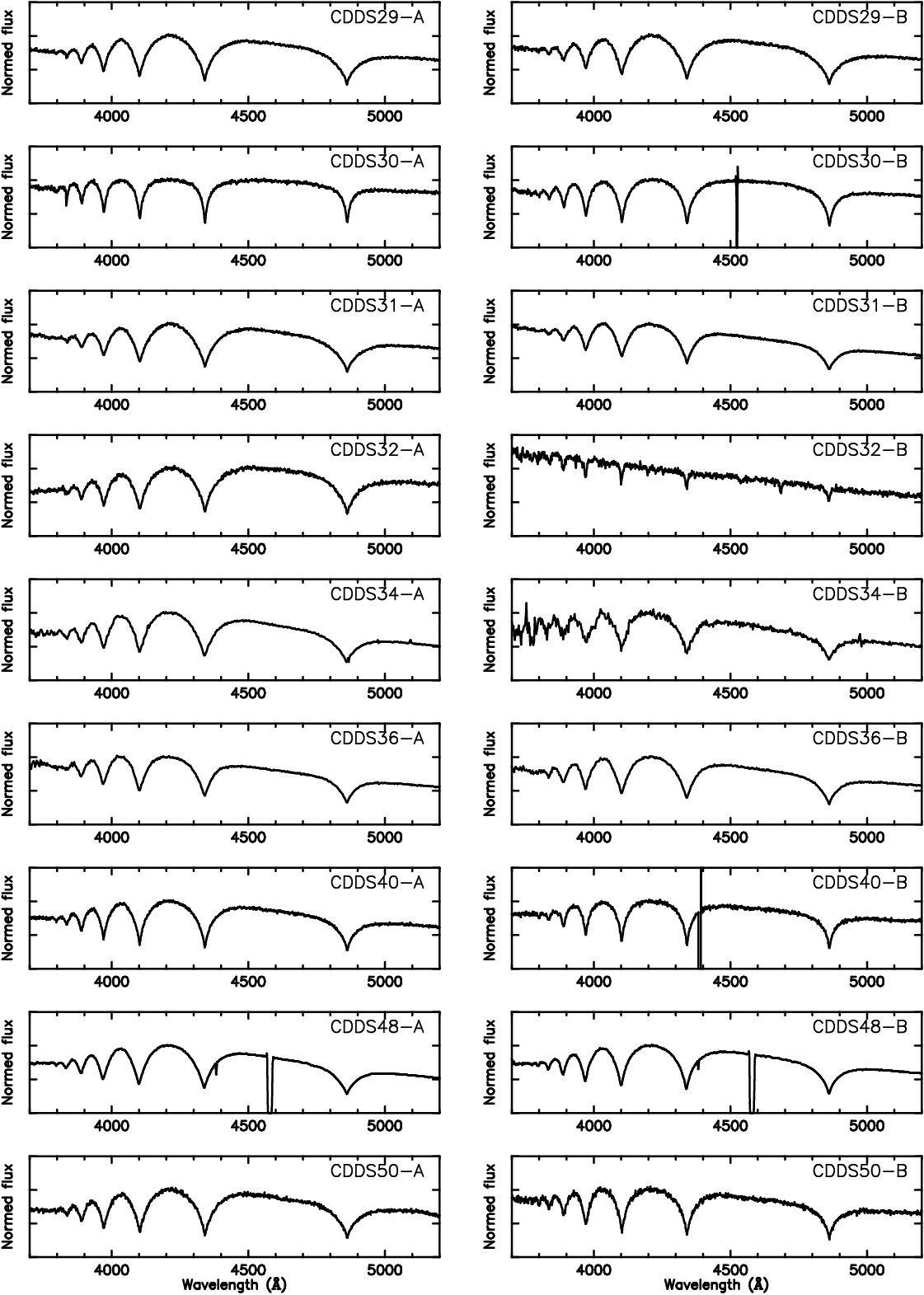}
\caption{Low resolution optical spectroscopy for the components of candidate binary systems in the range RA=12--24hr. These data have been normalised by dividing by the median flux in the interval $\lambda$=4180 -- 4220\AA.}
\label{specs2}
\end{figure*}

\section{The initial selection of candidate young, wide, double-degenerates from the SDSS DR7}
\label{selection}

An initial search for candidate, young, wide, white dwarf + white dwarf binaries was conducted using photometry
obtained from the SDSS DR7 database \citep[e.g.][]{baxter11}. The SDSS is a deep, wide-area, five band ($u$ (3551\AA),
$g$ (4686\AA), $r$ (6165\AA), $i$ (7481\AA), and $z$ (8931\AA)) imaging survey of the night sky that is supplemented by 
fiber spectroscopic follow-up of select sources (e.g. quasars). A comprehensive description of this impressively large 
project can be found in \cite{york00}, while full details of the DR7, which is of particular relevance to the present work, 
are provided by \cite{abazajian09}. In brief, the 7th data release includes imaging for an area of 11\,663 square degrees 
(a substantial proportion of which is centered on the northern Galactic cap) and catalogues 357 million unique sources 
down to 5$\sigma$ photometric limits at $u$, $g$, $r$ and $i$ of 22.3, 23.3, 23.1 and 22.3 mag., respectively. The 
imaging was acquired with a drift scan technique in seeing of better than 1.5 arcsec, so the median full width half-maximum 
of the point spread function (PSF) is approximately $1.4$ arcsec at $r$. The re-constructed SDSS images have pixels which 
are 0.396 arcsec on a side.

As is evident from Figure 1 of \cite{harris03} the photometric band passes of the SDSS and the colours which can be derived 
from them are rather effective at discriminating hot ($T_{\rm eff}$$>$8000--9000K), generally young, white dwarfs from the 
bulk of the field main sequence stars which dominate colour-magnitude and colour-colour diagrams. We selected from DR7 (using 
an SQL query) all point sources flagged as photometrically clean with $r\le20.0$, $u - g > -0.7$, $u - g \le 0.5$, $g - r >-0.7$,
$g - r \le 0.0$ and $r - i < 0.0$, which have another object satisfying these colour and magnitude criteria within 30 arcsec.
In drawing up these criteria we crudely appraised the likely contribution of chance stellar alignments to our sample of candidate 
binaries. We compared the cumulative number of observed pairings with separations of less than 60 arcsec to that predicted for a 
random on-sky distribution of objects by Equation 1 \citep[][]{struve52}, where $N$ is the number of sources satisfying the photometric 
selection criteria ($N$=36231) in area $A$ (square degrees, $A$=11663) and $\rho$ is the maximum projected separation (degrees),

\begin{eqnarray}
n(\le\rho)= N (N-1) \pi \rho^{2} / 2 A
\end{eqnarray}

On this basis we anticipate approximately 90\% of the pairings with separations of less than 30 arcsec to be physical systems 
(Figure~\ref{cumulate}), although we note our estimate neglects the variations of Galactic latitude within the sample and that a sizeable 
proportion of sources around $u - g=0.0-0.5$, $g - r=-0.2-0.0$ are likely to be quasars. Additionally, we considered that the atmospheres 
of cooler white dwarfs (ie. $T_{\rm eff}$$\simless$8000K) are significantly more difficult to model reliably due to the emergence of more 
physically complex sources of opacity \citep{koester10}. Moreover, by restricting our sample to relatively conservative faint magnitude limits 
we maintain the advantage that our sources can be followed-up spectroscopically on an 8m class telescope in reasonable integration times (e.g. 
about 15min in photometric conditions, or 2h in poor weather conditions).

The resulting 91 candidate systems were visually inspected using the SDSS finder chart tool to weed out a substantial number of spurious 
pairings attributable to blue point-like detections within nearby galaxies, which our contamination estimate above does not account for. To
identify further potential contaminants we also cross correlated our candidates against the SDSS DR7 quasar list of \cite{schneider10}. Our cleaned sample 
includes 53 candidate systems and recovers the three previously known double-degenerate binaries, PG\,0922+162 \citep{finley97}, HS\,2240+1234 
\citep{jordan98} and WD\,222301.62+220131.3 \citep{koester09} within the DR7 footprint. Details of all 53 candidates are listed in Table~\ref{allcands}.

\section[]{Spectroscopy}

\subsection{Observations and data reduction}
\label{survey}

While SDSS DR7 spectroscopy is available for objects in 18 of our systems (highlighted in italics in Table~\ref{allcands}), it often covers only 
one of the sources in a pairing \citep[e.g. CDDS\,40, ][]{kleinmann13}, or, where the component angular separation is less than the diameter of the SDSS 
fibres (3 arcsec), it provides a blend of the energy distributions of the two objects \citep[e.g. CDDS\,11,][]{dobbie13a, kepler13}. Therefore, to confirm or otherwise
the degenerate nature of both components of the candidate binaries in Table~\ref{allcands} and to determine their fundamental parameters, we have acquired
our own spatially resolved low-resolution optical spectroscopic observations with a range of facilities, as described below. As much of this spectroscopy was 
acquired in queue scheduling mode under less than optimum sky conditions, many of the systems we have followed-up are from the brighter half of our sample
and have angular separations that are greater than 2 arcsec. There is some minor overlap between our observations and the SDSS spectroscopy (CDDS\,8-A + B,
CDDS\,26-B, CDDS\,30-B and CDDS\,40-A) and the results of the \cite{kleinmann13} analyses of these DR7 datasets can serve as a useful check on our work.

For bona-fide degenerate objects, we have used our new spectroscopic data to determine the dominant elemental component of their atmospheres and whether 
or not they harbour a substantial magnetic field. In this work we have focused on the non-magnetic white dwarfs and have measured their effective temperatures
and surface gravities by comparing suitable absorption features within their spectra (e.g. H-Balmer lines) to the predictions of modern model atmospheres. 
Spectroscopically observed pairings with at least one strongly magnetic component are analysed elsewhere \citep[e.g. CDDS11, CDDS19, CDDS22, CDDS27, 
CDDS33, CDDS52, ][Dobbie et al. in prep.]{dobbie12a, dobbie13a}. The five different telescopes we have sourced spectroscopy from are: the William Herschel 
Telescope (WHT) and the Gran Telescopio Canarias (GTC), both located at the Roche de las Muchachos Observatory on La Palma, Gemini-North at the Mauna Kea
Observatory on Hawaii, Gemini-South at Cerro Pachon in Chile and the Very Large Telescope (VLT) Antu located at the European Southern Observatory's (ESO)
Cerro Paranal site in Chile. Details of the observations obtained with each facility are provided below:

\begin{itemize}

\item{Spectra of CDDS3, CDDS40 and CDDS50 were acquired in visitor mode with the WHT and the double-armed Intermediate dispersion Spectrograph and Imaging System 
(ISIS) on the nights of 2008 July 24--25 and 2011 September 5. These observations were conducted when the sky was clear, with seeing $\sim$0.6-0.9 arcsec. The spectrograph 
was configured with a 1.0 arcsec slit and the R300B ($\lambda$/$\delta\lambda$$\approx$1300) grating on the blue arm. The long exposures  necessary to reach good signal-to-noise
 were broken down into several integrations of typically 1800s (see Table 2),  and were acquired using the 2$\times$1 and 1$\times$1 binning modes of the $E2V$ CCD during the 2008 and 2011 runs, respectively. The data frames were debiased and flat fielded using the 
{\tt IRAF} procedure {\tt CCDPROC}. Cosmic ray hits were removed using the routine {\tt LACOS SPEC} \citep{vanDokkum01}. Subsequently, the spectra were extracted with the {\tt APEXTRACT}
package and wavelength calibrated by comparison with a CuAr+CuNe arc spectrum taken immediately before and after the science exposures. The removal of remaining instrument 
signature from the science data was undertaken using observations of the bright DC white dwarfs WD\,1918+386 and EG131.}

\item {Spectra of CDDS8, CDDS26, CDDS29, CDDS30, CDDS31 and CDDS32 were obtained in visitor mode with VLT Antu and the Focal Reducer and low dispersion Spectrograph 
(FORS2). A full description of the FORS2 instrument may be found on the ESO webpages\footnote{http://www.eso.org/instruments/fors2/}. These observations were conducted on the 
nights of 2010 February 6--7 and 2013 February 10--11.  The sky conditions were fair to good at the time of these observations. All data were acquired using the 2$\times$2 binning
mode of the $E2V$ CCD and the 600B+24 grism.  A 1.3 arcsec wide and a 1.0 arcsec wide slit were used for the 2010 and 2013 observations, respectively, providing notional spectral 
resolutions of $\lambda$/$\delta$$\lambda$$\sim$600 and $\lambda$/$\delta$$\lambda$$\sim$800. The data were reduced and extracted with {\tt IRAF} routines, as per the ISIS spectra, 
and wavelength calibrated by comparison with a He+HgCd arc spectrum obtained within a few hours of the science frames. Remaining instrument signature was removed from the science data 
using observations of the bright DC white dwarf LHS2333.}

\item {Spectra of CDDS6 and CDDS48 were obtained in service mode with the Gemini multi-object spectrograph (GMOS) mounted on the Gemini-North and Gemini-South telescopes during semesters
2009B (July 25) and 2010A (May 11), respectively. These observations were conducted when the sky conditions were relatively poor 
(image quality 85\% and cloud cover 90\%). The data were acquired using the 4$\times$4 binning mode of the EEV CCD, a 2.0 arcsec wide slit and the B600 grating tuned to a 
central wavelength of 4100\AA. The notional resolution of these spectra is $\lambda$/$\delta$$\lambda$$\sim$600. The data were reduced and extracted using routines in the 
{\tt GEMINI IRAF} software package. A wavelength solution for these data was obtained with a CuAr arc spectrum obtained within hours of the science frames, while residual instrument
signature in the science data was removed using observations of the bright DC white dwarfs WD\,1918+386 (North) and WD\,0000-345 (South).}

\item {Spectra of CDDS9, CDDS14, CDDS15, CDDS16, CDDS17, CDDS34 and CDDS36 were acquired in service mode with the GTC and the Optical System for Imaging and low Resolution 
Integrated Spectroscopy (OSIRIS) during semester 2013A. A detailed description of the OSIRIS instrument is provided on the GTC webpages
\footnote{http://www.gtc.iac.es/instruments/osiris/osiris.php}. 
Our observations were performed when sky conditions were less favourable (seeing$\approx$1.2--1.5 arcsec and spectroscopic transparency),   using the 2$\times$2 binning mode of the EEV CCD. With the R1000B grating, inconjunction with a 1.2 arcsec wide slit, we 
achieved a notional spectral resolution of $\lambda$/$\delta\lambda$$\approx$500. The data were reduced and extracted with {\tt IRAF} routines as per the ISIS spectra. Wavelength 
calibration was performed by comparing these to a Hg + Ne arc spectrum acquired within a few hours of the science frames. Particular care was taken to include the 3650.1\AA\ HgI line at 
the very blue limit of our spectral coverage in our arc wavelength solution. Remaining instrument signature from the science data was removed using observations of the bright DC white 
dwarf WD\,1918+386.}

\end{itemize}

The spectroscopic observations are summarised in  Table 2. We noted above that three of our candidate double-degenerate systems have been the subjects 
of previous spectroscopic studies, PG\,0922+162 \citep{finley97}, HS\,2220+2146 \citep{koester09} and HS\,2240+1234 \citep{jordan98}. Throughout the rest of this 
work we adopt the parameters measured from a high-spectral resolution \cite{koester09} investigation of the components of these three pairings. We re-affirm their results for HS\,2220+2146 from our measurements of a low-resolution Gemini-N/GMOS spectrum ($T_{\rm eff}$=19020$\pm$438 K, log~$g$=8.37$\pm$0.07 and $T_{\rm eff}$=13950$\pm$321 K, log~$g$=8.07$\pm$0.07 for SDSS\,J222301.62+220131.3 and SDSS\,J222301.72+220124.9, respectively).

\section{Data analysis}

\subsection{White dwarfs, effective temperatures, surface gravities and distances}
\label{disto}

Our optical spectroscopy (Figure~\ref{specs1}) reveals through the  presence of broad H-Balmer line series, that the overwhelming majority 
of the objects we have followed-up are white dwarfs with hydrogen-rich atmospheres (DAs). As discussed in companion papers, all the non-magnetic 
white dwarfs in six additional spectroscopically observed pairings harbouring a strongly magnetic component, also appear to have hydrogen dominated atmospheres 
\citep[e.g.][]{dobbie12a,dobbie13a}. The spectrum of one object in our sample, CDDS15-A (SDSS\,J084952.87+471249.4), features strong, pressure broadened, HeI lines, 
consistent with if being a degenerate with a helium dominated atmosphere (DB). The energy distribution of CDDS32-B (SDSS\,J135713.14-065913.7) appears to be 
that of an evolved subdwarf star (sdO) which is likely to be located at a distance of several kiloparsec (compared to only several hundred parsec for the white 
dwarf CDDS32-A). Consequently, the CDDS32 pairing is not considered further. 

We have measured the effective temperatures and surface gravities of the DAs by comparing the observed lines, H-8 to H-$\beta$, to a grid of synthetic profiles  
based on recent versions of the plane-parallel, hydrostatic, local thermodynamic equilibrium (LTE) atmosphere and spectral synthesis codes {\tt ATM} and {\tt SYN} 
\citep[e.g.][]{koester10}. These models include an updated physical treatment of Stark broadening of {\tt HI} lines \citep{tremblay09}. The effective 
temperature and surface gravity of the DB (CDDS15-A) has been measured by comparing the normalised observed energy distribution within the wavelength range 3750-5150\AA,
to a grid of similarly normalised synthetic spectra. These were also generated using {\tt ATM} and {\tt SYN}, in this case tuned for the treatment of helium rich atmospheres.
Our model fitting was undertaken with the spectral analysis package {\tt XSPEC} \citep{shafer91} as described in previous work \citep[e.g.][]{dobbie09a}.  The errors in our parameter determinations have been calculated by stepping that in question away from its optimum value until the difference between the two values of the fit statistic ($\Delta\chi^{2}$), corresponds to 1$\sigma$ for a given number of free model parameters \citep[e.g.][]{lampton76}. However, as these are only formal (internal) estimates of the errors on effective temperature and surface gravity, and may underestimate the true uncertainties, we have assumed their magnitudes to be at least 2.3\% and 0.07dex, respectively, for the DA stars \citep{napiwotzki99} and at least 2.3\% and 0.05dex, respectively, for the DB type objects \citep{bergeron11}. In cases where the fitting provided two solutions for the effective temperature of a star, we compared its observed photometry to synthetic colours \citep[e.g.][]{bergeron95} to break the degeneracy.

\begin{figure}
\includegraphics[angle=0,width=9cm]{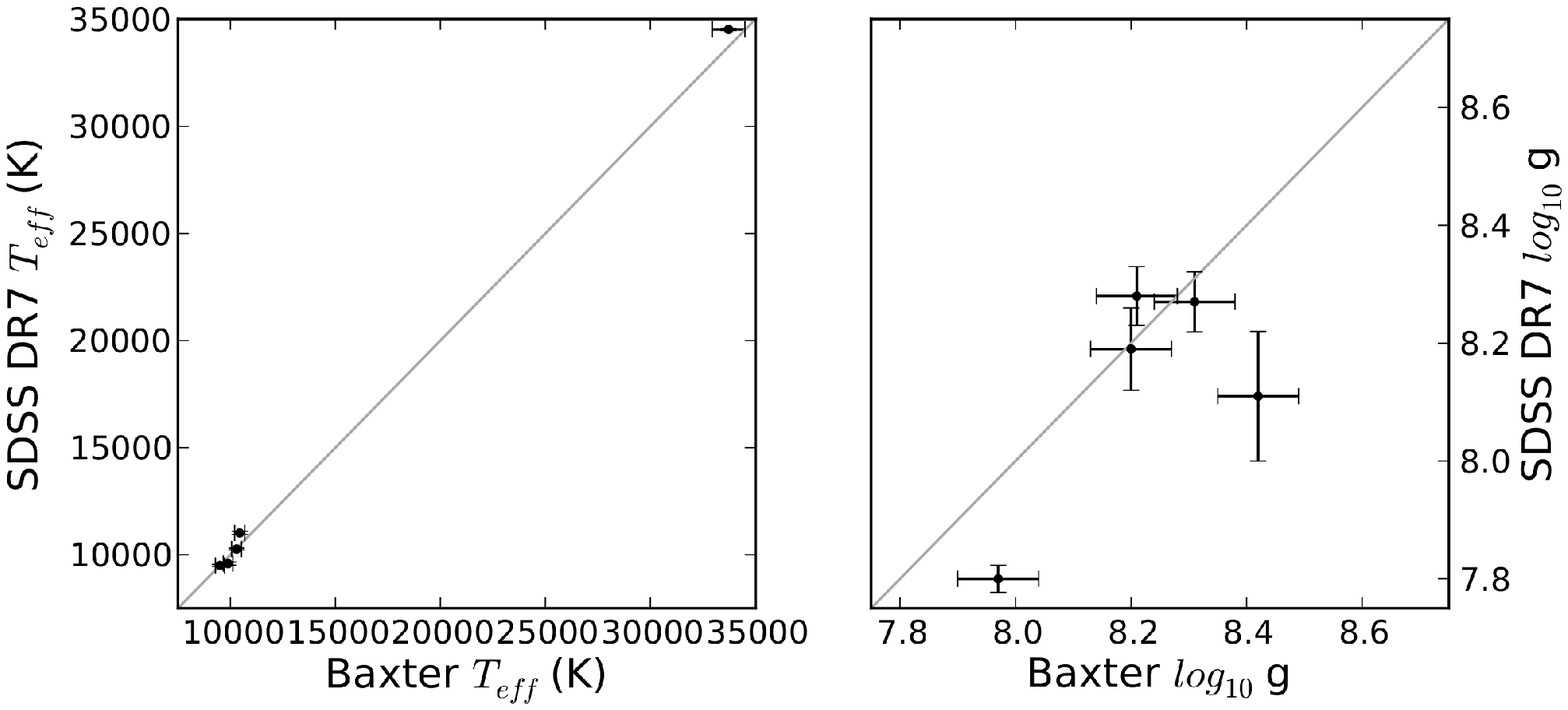}
\caption{Plot of our estimates of effective temperature and surface gravity against the SDSS DR7 spectroscopic based measurements of {\protect \cite{kleinmann13}} for five 
white dwarfs in common to our spectroscopic sample and the DR7 white dwarf catalogue.}
\label{teffteff}
\end{figure}

The results of our analysis procedure are listed in Table~\ref{temps}. As can be seen from Figure~\ref{teffteff}, our measurements of these parameters for the 
five objects that are common to both the DR7 white dwarf catalogue and our spectroscopic follow-up are generally accordant with those of \cite{kleinmann13}. The surface 
gravity estimates for CDDS\,8-A and CDDS\,40-B deviate at around 2$\sigma$ from the one-to-one relation. It is conceivable that these differences are statistical in nature 
but, considering our observations for both these objects have provided data of excellent signal-to-noise for the gravity sensitive spectral features, systematic error could 
also be the cause. The former object is the hottest white dwarf in our spectroscopic sample and has relatively weak high order Balmer lines in the wavelength regime 
($\lambda<4000$\AA) where the response of the SDSS spectrographs changes shape (decreases) rapidly. The effective temperature we measure for CDDS\,8-B is also lower by around
500K than the \cite{kleinmann13} determination, yet both the SDSS data and our spectrum for this object provide strong signal-to-noise coverage of the potent temperature 
diagnostic Balmer lines e.g. H$\gamma$ and H$\beta$. We discuss this object in more detail in Section~\ref{zzceti}, but for now note that it is a pulsating white dwarf. 

\begin{figure}
\includegraphics[angle=0,width=\linewidth]{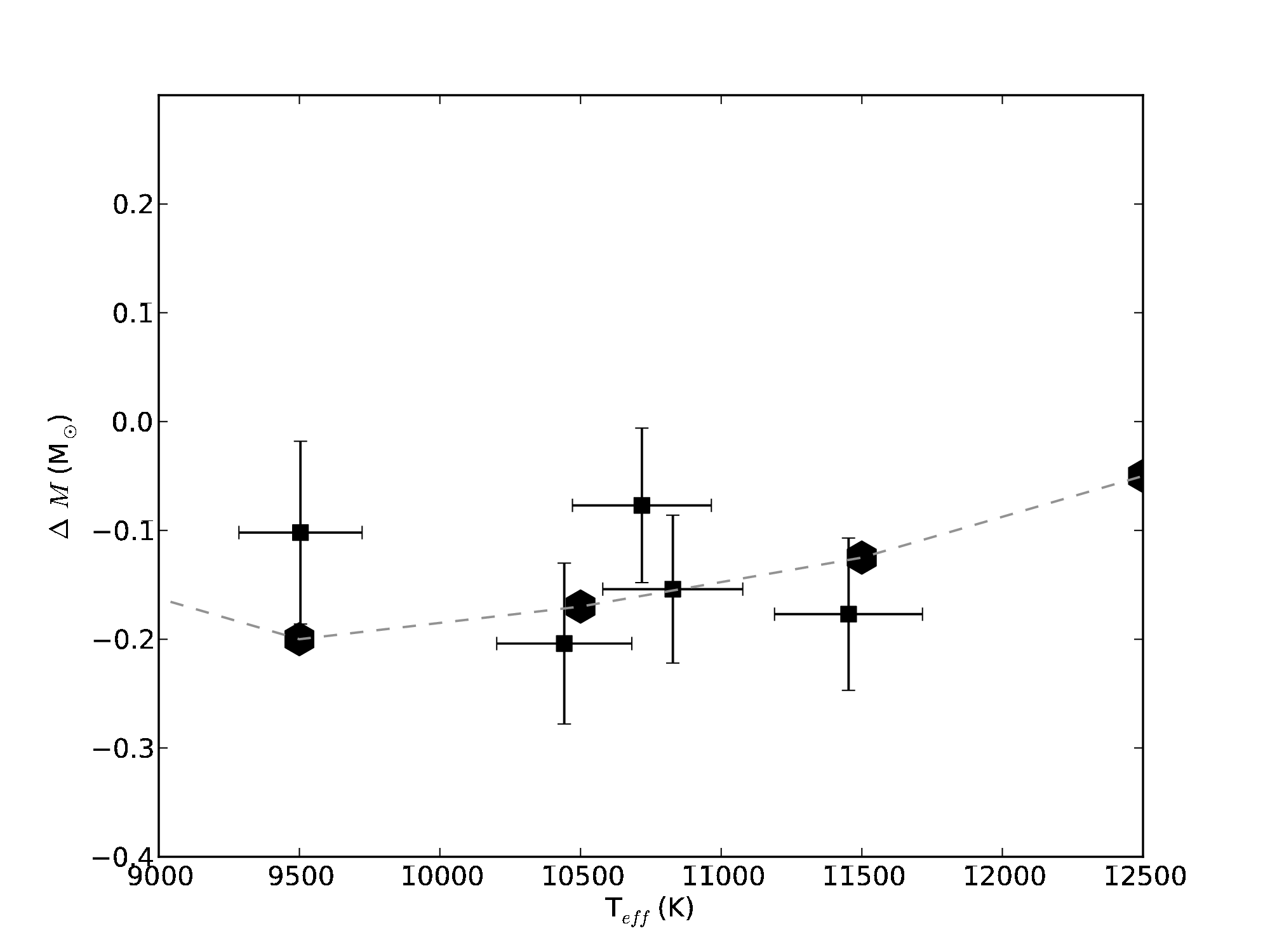}
\caption{ Plot of the departures between the wholly spectroscopically derived mass and that derived from the spectroscopic effective temperature and the SDSS photometry 
under the assumption that the components lie at the same distance, for five white dwarfs with $T_{\rm eff}$$<$12500K in wide double-degenerate
pairings where the companion stars have $T_{\rm eff}$$>$12500K and the components have highly significant and consistent proper motions (filled squares, from left to right, CDDS26-B,
CDDS8-B, CDDS50-B, CDDS3-A and CDDS48-A). The increase in the mean spectroscopic mass below $T_{\rm eff}$$<$12500K, with respect to the value at $T_{\rm eff}$$>$13000K, found by 
{\protect \cite{tremblay11}} in their study of the SDSS DR4 white dwarf sample is also shown (filled hexagons and dashed line).}
\label{massoff}
\end{figure}

\begin{figure*}
\includegraphics[angle=0,width=12cm]{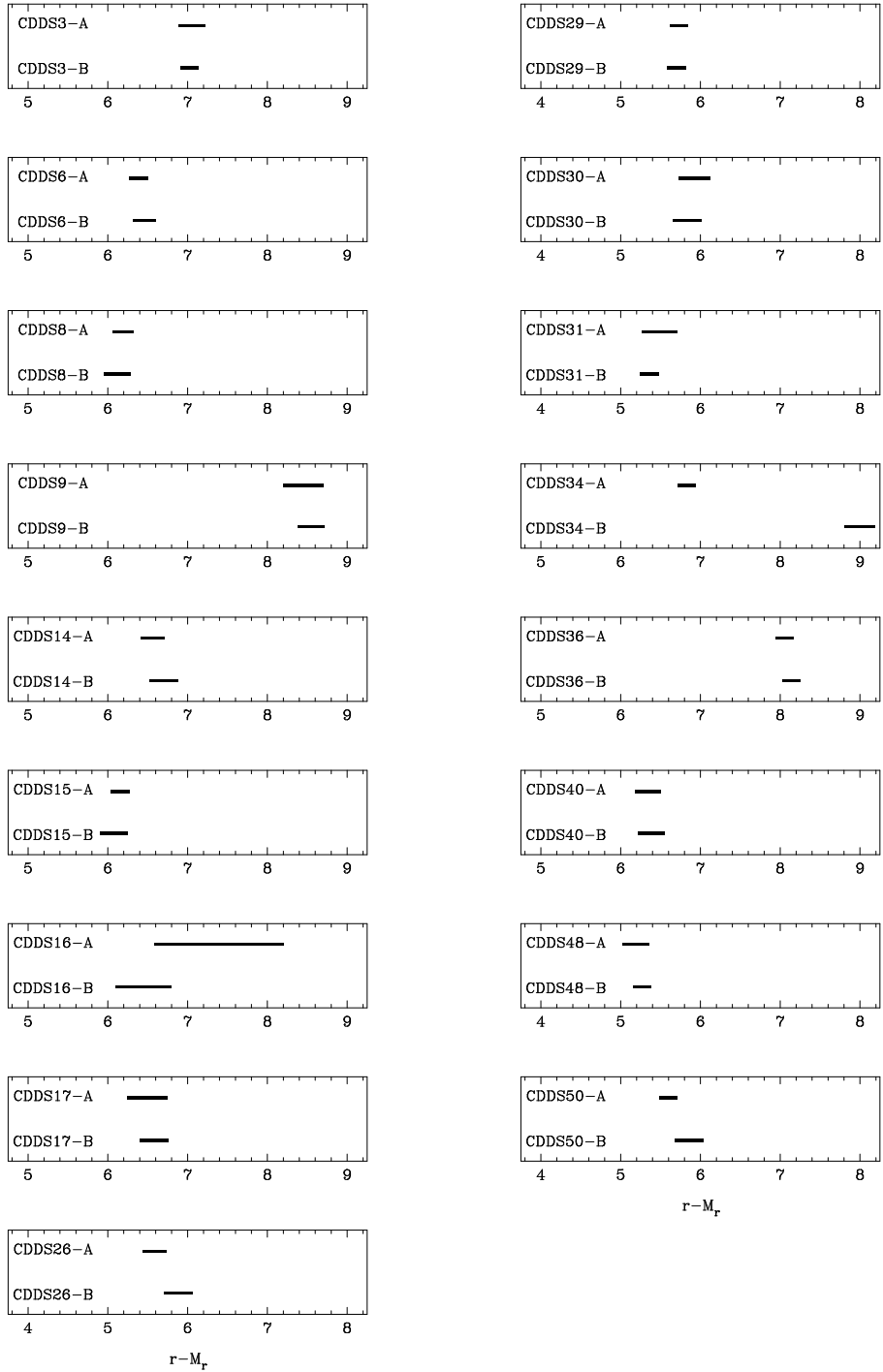}
\caption{The derived distance modulii of the white dwarfs in our candidate binaries. The components of all but one system have 
consistent distances, as expected for physical binary systems.}
\label{dmod}
\end{figure*}

Subsequently we have derived the absolute $r$ magnitudes of each white dwarf by interpolating (using cubic-splines) within grids of modern synthetic photometry
\footnote{http://www.astro.umontreal.ca/~bergeron/CoolingModels} to the measured effective temperature and surface gravity. The He-rich grids, relevant to CDDS15-A, originate 
from the work of \cite{bergeron11}. The H-rich grids, appropriate to the DAs, are based on the work of \cite{bergeron95} but revised to include updates from \cite{holberg06},
\cite{kowalski06} and \cite{tremblay11}. Spectroscopically derived effective temperatures and surface gravities are known to lead to systematic overestimates of DA white dwarf 
masses at $T_{\rm eff}$$<$12500K \citep[e.g.][]{kepler07,bergeron95}.  We have adjusted the derived parameters of our affected objects by subtracting mass offsets, which have been
determined from the data points shown in Figure 19 of \cite{tremblay11}. These data points have been verified to provide a fair representation of the systematic overestimate 
through examining them with respect to the departures between the two mass estimates we have made for the subset of our cool degenerates that are paired with a white dwarf 
with $T_{\rm eff}$$>$12500K, ie. the wholly spectroscopic mass estimate and the mass derived from the spectroscopic effective temperature and the SDSS photometry under the 
assumption that the components lie at the same distance (Figure~\ref{massoff} and next section).
Finally, to help appraise the physical reality of our pairs we have calculated the distance moduli of the white dwarfs, neglecting foreground extinction (Table~\ref{wdmass} and 
Figure~\ref{dmod}). Dust maps suggest this is very low along the majority of these Galactic lines of sight \citep{schlegel98}. Moreover, our conclusions are primarily sensitive to 
differential reddening and given the angular proximity of our targets, for physical systems this is likely to be very small. 

\begin{table*}
\begin{minipage}{180mm}
\begin{center}
\caption{Our effective temperature and surface gravity measurements for the components of the wide white dwarf + white dwarf candidate binary systems. The predicted 
absolute $r$ magnitudes, distance modulii, masses and proper motions for these objects are also listed. Where deemed necessary, parameters have been modified to account for the
spectroscopic overestimate of mass which occurs at $T_{\rm eff}$$<$ 12500K (column 4; see text for further details). An assessment of the physical nature of each pairing has been
made by considering a Bayes factor estimated from the distances and the astrometry of the components.}
\label{wdmass}
\begin{tabular}{clcccccccc}
\hline
ID & \multicolumn{1}{c}{$T_{\rm eff}$} & log~$g$ & $\Delta$$M$  & M$_{r}$ & $r$-M$_{r}$ & $M$  &  $\mu_{\alpha}\cos\delta$,$\mu_{\delta}$  & $\frac{P(data \mid C)}{P(data \mid F)}$ & DD$?$ \\ 
& \multicolumn{1}{c}{/K} & /$\log_{10} (cm\,s^{-2})$ & /M$_{\odot}$  & /mag. & /mag. & /M$_{\odot}$  &  /mas~yr$^{-1}$ &  & \\ 

\hline

\multicolumn{1}{l}{CDDS3-A} & 10828$\pm$249 & 8.16$\pm$0.07 & -0.16 &11.87$^{+0.16}_{-0.16}$ & 7.05$\pm$0.17 & 0.54$\pm$0.06 & 4.7$\pm$4.4,~-41.6$\pm$4.4 & \multirow{2}{*}{345} & \multirow{2}{*}{\bf\prob}\\ 
\multicolumn{1}{l}{CDDS3-B} & 19116$\pm$440 & 7.94$\pm$0.07 & -- &10.96$^{+0.12}_{-0.12}$ & 7.02$\pm$0.11 & 0.59$\pm$0.04 & 8.6$\pm$4.4,~ -41.8$\pm$4.4\\ \\

\multicolumn{1}{l}{CDDS6-A} & 24788$\pm$570 & 8.06$\pm$0.07 & -- &10.69$^{+0.11}_{-0.12}$ & 6.39$\pm$0.12 & 0.67$\pm$0.04 & -50.9$\pm$18.3,~-2.6$\pm$10.7 & \multirow{2}{*}{91} & \multirow{2}{*}{\bf\prob}\\
\multicolumn{1}{l}{CDDS6-B} & 18502$\pm$426 & 8.06$\pm$0.07 & -- &11.19$^{+0.11}_{-0.11}$ & 6.46$\pm$0.14 & 0.65$\pm$0.04 & -34.7$\pm$18.3,~3.6$\pm$10.7\\ \\


\multicolumn{1}{l}{CDDS8-A} & 33725$\pm$776 & 7.97$\pm$0.07 & -- & 9.90$^{+0.13}_{-0.12}$ & 6.19$\pm$0.13 & 0.64$\pm$0.04 & -31.2$\pm$3.2,~-23.6$\pm$3.2 & \multirow{2}{*}{1395} & \multirow{2}{*}{\bf\prob} \\
\multicolumn{1}{l}{CDDS8-B} & 10442$\pm$240 & 8.31$\pm$0.07 & -0.17 & 12.18$^{+0.17}_{-0.16}$ & 6.12$\pm$0.17 & 0.63$\pm$0.06 & -28.4$\pm$3.2,~-23.9$\pm$3.2\\ \\


\multicolumn{1}{l}{CDDS9-A} & 11636$\pm$410 & 7.95$\pm$0.13 & -0.11 & 11.51$^{+0.24}_{-0.25}$ & 8.45$\pm$0.25 & 0.47$\pm$0.04 & -4.0$\pm$3.9,~-9.8$\pm$3.9 & \multirow{2}{*}{11} & \multirow{2}{*}{\prob}\\ 
\multicolumn{1}{l}{CDDS9-B} & 16410$\pm$524 & 7.96$\pm$0.11 & -- & 11.25$^{+0.17}_{-0.17}$ & 8.55$\pm$0.17  & 0.59$\pm$0.05 & 2.6$\pm$4.1,~-14.2$\pm$3.9\\ \\


\multicolumn{1}{l}{CDDS14-A} & 12185$\pm$280 & 8.11$\pm$0.07 & -0.07 &11.76$^{+0.15}_{-0.14}$ & 6.56$\pm$0.15 & 0.60$\pm$0.06 & 4.0$\pm$3.2,~-2.1$\pm$3.2 & \multirow{2}{*}{130} & \multirow{2}{*}{\bf\prob}\\
\multicolumn{1}{l}{CDDS14-B} & 10140$\pm$233 & 8.23$\pm$0.08 & -0.18 &12.12$^{+0.18}_{-0.18}$ & 6.70$\pm$0.18 & 0.56$\pm$0.06 & 5.6$\pm$3.2,~-0.4$\pm$3.2\\ \\

\multicolumn{1}{l}{CDDS15-A} & 16992$\pm$391 & 7.84$\pm$0.07 & -- &10.92$^{+0.11}_{-0.11}$ & 6.16$\pm$0.12 & 0.51$\pm$0.04 & -88.0$\pm$3.0,~-76.9$\pm$3.0 & \multirow{2}{*}{$>$10000} & \multirow{2}{*}{\bf\prob}\\
\multicolumn{1}{l}{CDDS15-B} & 11127$\pm$256 & 8.08$\pm$0.07 & -0.14 &11.72$^{+0.16}_{-0.16}$ & 6.07$\pm$0.17 & 0.51$\pm$0.06 & -82.3$\pm$3.0,~-76.5$\pm$3.0\\ \\

\multicolumn{1}{l}{CDDS16-A} & 11032$\pm$872 & 7.94$\pm$0.39 & -0.15 & 11.48$^{+0.70}_{-0.90}$ & 7.39$\pm$0.81 & 0.42$\pm$0.22 & 3.0$\pm$4.1,~3.4$\pm$4.1 & \multirow{2}{*}{10} & \multirow{2}{*}{\prob}\\
\multicolumn{1}{l}{CDDS16-B} & 11837$\pm$824 & 8.21$\pm$0.18 & -0.10 &11.89$^{+0.35}_{-0.40}$ & 6.45$\pm$0.35 & 0.63$\pm$0.12 & -0.8$\pm$4.1,~-0.3$\pm$4.1\\ \\

\multicolumn{1}{l}{CDDS17-A} & 10074$\pm$232 & 8.46$\pm$0.13 & -0.18 & 12.51$^{+0.25}_{-0.24}$ & 6.50$\pm$0.25 & 0.71$\pm$0.09 & 15.3$\pm$2.9,~-15.1$\pm$2.9 & \multirow{2}{*}{119} & \multirow{2}{*}{\bf\prob}\\
\multicolumn{1}{l}{CDDS17-B} & 11083$\pm$254 & 8.23$\pm$0.08 & -0.15 &11.95$^{+0.17}_{-0.17}$ & 6.58$\pm$0.18 & 0.60$\pm$0.06 & 14.3$\pm$2.9,~-9.2$\pm$2.9\\ \\

\multicolumn{1}{l}{CDDS26-A} & 16401$\pm$377 & 8.94$\pm$0.07 & -- & 12.93$^{+0.15}_{-0.14}$ & 5.59$\pm$0.15 & 1.17$\pm$0.03 & -19.9$\pm$3.0,~-42.8$\pm$3.0 & \multirow{2}{*}{1266} & \multirow{2}{*}{\bf\prob}\\ 
\multicolumn{1}{l}{CDDS26-B} & 9504$\pm$219 & 8.20$\pm$0.07 & -0.20 & 12.24$^{+0.18}_{-0.18}$ & 5.88$\pm$0.18 & 0.52$\pm$0.06 & -21.1$\pm$3.0,~-45.2$\pm$3.0\\ \\       

\multicolumn{1}{l}{CDDS29-A} & 12758$\pm$293 & 8.18$\pm$0.07 & -- &11.96$^{+0.11}_{-0.10}$ & 5.73$\pm$0.11 & 0.72$\pm$0.04 & -31.6$\pm$2.9,~26.4$\pm$2.9 & \multirow{2}{*}{1992} & \multirow{2}{*}{\bf\prob}\\
\multicolumn{1}{l}{CDDS29-B} & 13072$\pm$301 & 8.41$\pm$0.07 & -- &12.28$^{+0.12}_{-0.11}$ & 5.70$\pm$0.12 & 0.87$\pm$0.05 & -26.8$\pm$2.9,~27.8$\pm$2.9\\ \\


\multicolumn{1}{l}{CDDS30-A$^{*}$} & 9203$\pm$212 & 8.14$\pm$0.07 & -0.18 & 12.30$^{+0.17}_{-0.18}$ & 5.92$\pm$0.20 & 0.50$\pm$0.06 & -45.3$\pm$14.9,~33.5$\pm$14.8 & \multirow{2}{*}{125} & \multirow{2}{*}{\bf\prob}\\
\multicolumn{1}{l}{CDDS30-B} & 10293$\pm$237 & 8.21$\pm$0.07 & -0.18 & 12.04$^{+0.17}_{-0.18}$ & 5.84$\pm$0.18 & 0.55$\pm$0.06 & -43.1$\pm$14.9,~32.5$\pm$14.8\\ \\

\multicolumn{1}{l}{CDDS31-A} & 14037$\pm$323 & 8.45$\pm$0.07 & -- & 12.25$^{+0.12}_{-0.12}$ & 5.49$\pm$0.22 & 0.89$\pm$0.05 & -26.3$\pm$27.3,~20.0$\pm$12.7 & \multirow{2}{*}{54} & \multirow{2}{*}{\bf\prob}\\
\multicolumn{1}{l}{CDDS31-B} & 20000$\pm$460 & 8.19$\pm$0.07 & -- &11.27$^{+0.12}_{-0.12}$ & 5.36$\pm$0.12 & 0.74$\pm$0.04 & -50.8$\pm$27.3,~22.8$\pm$12.7\\ \\




\multicolumn{1}{l}{CDDS34-A} & 14100$\pm$398 & 7.86$\pm$0.07 & -- &11.37$^{+0.11}_{-0.11}$ & 6.83$\pm$0.11 & 0.53$\pm$0.04 & 0.2$\pm$3.5,~-22.2$\pm$3.5 &\multirow{2}{*}{$<<1$} & \multirow{2}{*}{\nae}\\
\multicolumn{1}{l}{CDDS34-B} & 18261$\pm$598 & 7.90$\pm$0.12 & -- &10.98$^{+0.18}_{-0.18}$ & 9.00$\pm$0.19 & 0.56$\pm$0.07 & 0.1$\pm$3.5,~-0.2$\pm$3.5\\ \\

\multicolumn{1}{l}{CDDS36-A} & 18769$\pm$432 & 7.94$\pm$0.07 & -- &10.99$^{+0.11}_{-0.11}$ & 8.06$\pm$0.11 & 0.59$\pm$0.04 & -5.7$\pm$4.2,~-12.5$\pm$4.2 & \multirow{2}{*}{72} & \multirow{2}{*}{\bf\prob}\\
\multicolumn{1}{l}{CDDS36-B} & 16542$\pm$380 & 7.91$\pm$0.07 & -- &11.16$^{+0.11}_{-0.11}$ & 8.14$\pm$0.11 & 0.57$\pm$0.04 & -7.7$\pm$4.2,~-7.7$\pm$4.2\\ \\

\multicolumn{1}{l}{CDDS40-A} & 11207$\pm$258 & 8.22$\pm$0.07 & -0.14 &11.93$^{+0.16}_{-0.16}$ & 6.34$\pm$0.16 & 0.60$\pm$0.06 & 2.5$\pm$3.1,~-50.8$\pm$3.1 & \multirow{2}{*}{2418} & \multirow{2}{*}{\bf\prob}\\
\multicolumn{1}{l}{CDDS40-B} & 9888$\pm$227 & 8.42$\pm$0.07 & -0.19 &12.48$^{+0.17}_{-0.16}$ & 6.38$\pm$0.17 & 0.68$\pm$0.06 & 1.3$\pm$3.1,~-50.6$\pm$3.1\\ \\

\multicolumn{1}{l}{CDDS48-A} & 11453$\pm$263 & 8.35$\pm$0.07 & -0.13  &12.11$^{+0.16}_{-0.15}$ & 5.19$\pm$0.17 & 0.69$\pm$0.06 & -2.3$\pm$8.4,~-29.2$\pm$7.4 & \multirow{2}{*}{2530} & \multirow{2}{*}{\bf\prob}\\
\multicolumn{1}{l}{CDDS48-B} & 15834$\pm$364 & 8.02$\pm$0.07 & -- &11.40$^{+0.11}_{-0.11}$ & 5.27$\pm$0.11 & 0.63$\pm$0.04 & -0.0$\pm$8.4,~-28.0$\pm$7.4\\ \\

\multicolumn{1}{l}{CDDS50-A} & 12739$\pm$293 & 8.04$\pm$0.07 &  -- &11.76$^{+0.10}_{-0.10}$ & 5.60$\pm$0.11 & 0.63$\pm$0.04 & 38.3$\pm$10.8,~-0.9$\pm$8.8 & \multirow{2}{*}{126} & \multirow{2}{*}{\bf\prob}\\
\multicolumn{1}{l}{CDDS50-B} & 10718$\pm$247 & 8.26$\pm$0.07 & -0.16 &12.06$^{+0.16}_{-0.16}$ & 5.86$\pm$0.18 & 0.60$\pm$0.06 & 32.8$\pm$10.8,~-0.9$\pm$8.8\\ \\

\hline

\label{temps}
\end{tabular}
\end{center}
$^{\dagger}$ Proper motion measurements are obtained with respect to nearby stars \\
$^{*}$ Possible triple system, dK + WD + WD
\end{minipage}
\end{table*}

\begin{table*}
\begin{minipage}{180mm}
\begin{center}
\caption{Effective temperatures and surface gravities obtained from {\protect  \cite{koester09}} for the components of three previously identified wide 
DA + DA binary systems. We also list the theoretical absolute $r$ magnitudes, the distance modulii, the masses and the proper motions for these white 
dwarfs.}
\label{wdmass_lit}
\begin{tabular}{clcccccccc}
\hline
ID & \multicolumn{1}{c}{$T_{\rm eff}$} & log~$g$ & $\Delta$$M$  & M$_{r}$ & $r$-M$_{r}$ & $M$  &  $\mu_{\alpha}\cos\delta$,$\mu_{\delta}$  & $\frac{P(data \mid C)}{P(data \mid F)}$ & DD$?$ \\ 
& \multicolumn{1}{c}{/K} & /$\log_{10} (cm\,s^{-2})$ & /M$_{\odot}$  & /mag. & /mag. & /M$_{\odot}$  &  /mas~yr$^{-1}$ &  & \\ 
\hline


\multicolumn{1}{l}{CDDS18-A} & 23537$\pm$541 & 8.23$\pm$0.07 &-- & 11.05$^{+0.12}_{-0.12}$ & 5.50$\pm$0.12 & 0.77$\pm$0.04 & -49.9$\pm$9.9,~-30.3$\pm$9.9 & \multirow{2}{*}{61} & \multirow{2}{*}{\bf\prob}\\ 
\multicolumn{1}{l}{CDDS18-B} & 25783$\pm$593 & 9.04$\pm$0.07 &-- &12.39$^{+0.16}_{-0.16}$ & 5.12$\pm$0.18 & 1.22$\pm$0.03 & -33.5$\pm$9.9,~-34.0$\pm$9.9\\ \\

\multicolumn{1}{l}{CDDS49-A} & 14601$\pm$336 & 8.08$\pm$0.07 &-- &11.62$^{+0.11}_{-0.12}$ & 4.58$\pm$0.12 & 0.66$\pm$0.04 & 53.0$\pm$3.6,~-71.1$\pm$3.6 & \multirow{2}{*}{$>>$10000} & \multirow{2}{*}{\bf\prob}\\
\multicolumn{1}{l}{CDDS49-B} & 18743$\pm$431 & 8.24$\pm$0.07 &-- &11.45$^{+0.12}_{-0.12}$ & 4.46$\pm$0.12 & 0.77$\pm$0.04 & 54.0$\pm$3.6,~-75.5$\pm$3.6\\ \\

\multicolumn{1}{l}{CDDS51-A} & 15636$\pm$360 & 7.86$\pm$0.07 &-- &11.19$^{+0.11}_{-0.10}$ & 6.24$\pm$0.11 & 0.54$\pm$0.04 & -32.8$\pm$3.1,~-69.7$\pm$3.1 & \multirow{2}{*}{$>$10000} & \multirow{2}{*}{\bf\prob}\\
\multicolumn{1}{l}{CDDS51-B} & 13935$\pm$321 & 7.99$\pm$0.07 &-- &11.57$^{+0.11}_{-0.11}$ & 6.35$\pm$0.13 & 0.60$\pm$0.04 & -31.3$\pm$3.1,~-74.4$\pm$3.1\\ \\
                   
\hline
       
\label{wdmass_lit}
\end{tabular}
\end{center}
\end{minipage}
\end{table*}

\subsection{Astrometry}
\label{astrom}

To assess further the physical reality of our candidate binaries, we have performed astrometry on their components. In the majority of cases we have conducted these measurements
with the positions of sources from the full United States Naval Observatory B1.0 (USNO-B) catalogue \citep{monet03} and from the SDSS. We have followed a methodology similar to that
adopted by \cite{kraus07d} and the PPMXL survey \citep{roeser10} to shift the former onto the International Celestial Reference System (ICRS). 
In brief, we determined the tangent plane coordinates, $\xi$ and $\eta$, corresponding to each photographic epoch, for all sources within several degrees of our pairs from their
catalogued year 2000.0 positions, their tangent plane proper motions, their fit residuals and the dates of their photographic imaging. We next calculated for each object 
the celestial coordinates at the mean USNO epoch and at all photographic epochs (J2000.0) with the catalogued positions, the tangent plane coordinates for epoch 2000.0 
and routines in {\tt STARLINK SLALIB}. Subsequently, we cross-correlated sources brighter than $R2$=16.5 mag. with objects listed in the Fourth United States Naval Observatory 
CCD Astrograph Catalog \citep[UCAC4][]{zacharias13} that have low proper motions ($\mu$$<$5mas~yr$^{-1}$) and small proper motion uncertainties ($\Delta$$\mu$$<$5mas~yr$^{-1}$).
We derived the ICRS positions of the matched objects at each photographic epoch and finally, estimated any offsets between these and the USNO-B co-ordinates using a 
maximum-likelihood technique. 

We calculated absolute proper motions by performing a least-squares linear fit in each axis to the USNO-B co-ordinates on the ICRS and the SDSS positions as a function of date, 
weighting according to crude uncertainty estimates of 50 and 230 mas for SDSS and photographic epochs, respectively \citep[e.g.][]{kraus07d,roeser10}. In this final step we used 
only astrometry from photographic images where a visual inspection revealed the components to be clearly resolved. In cases where there was mild blending we relied on 
{\tt SEXTRACTOR} \citep[][]{bertin96} and a Mexican hat filter for a more robust discrimination of the components and their photocenters. However, several of our pairings were found to be sufficiently
blended in all photographic epochs that we had to adopt a different approach. In these cases we generally relied on the SDSS DR7 $r$ band co-ordinates as a first epoch and
positions from additional electronic imaging for the second epoch. 
We obtained $i$ band frames for CDDS6A and CDDS31 on 2011 February 12th using the Isaac Newton Telescope and Wide Field Camera. We extracted positions from $J$ band United Kingdon Infrared telescope Deep Sky Survey \citep[UKIDSS, ][]{lawrence07,casali07,hewett06,hodgkin09,hambly08} (2010 February 22nd) and $Y$ band Visible and Infrared Survey Telescope for Astronomy \citep[VISTA, ][]{mcmahon12} (2012 September 9th) imaging for CDDS30 and CDDS48, respectively. Positions for the components of CDDS50 were obtained from two epochs of SDSS $r$ band imaging separated by approximately 5 years (2004 September 17th and 2009 October 17th).  We employed routines in the {\tt STARLINK SLALIB} library to construct six co-efficient linear transforms between the two sets of coordinates for each putative system, iteratively clipping 3$\sigma$ outliers from the fits \citep[e.g.][]{dobbie02c}. Subsequently we determined relative proper motions by taking the differences between the observed and calculated locations of candidates in the second epoch imaging and dividing by the time baseline between the two observations. Astrometric measurements for the components of all spectroscopically observed pairs are displayed in Table ~\ref{wdmass}.

\subsection{Physical system or chance alignment$?$}

A casual inspection of Figure~\ref{dmod} reveals that our distance modulii estimates for the components in most pairings overlap with one and other. Moreover, the white 
dwarfs in 10 out of the 17 new candidate binaries have proper motions that are both significant ($\mu$ / $\delta\mu$ $\simgreat$3) and internally consistent, supporting the notion
that a sizeable majority of our pairs are physical systems. Only one system, CDDS\,34, displays a large and statistically significant (greater than 5$\sigma$) disparity between the 
distance modulii of its components. Our astrometric measurements also appear strongly discordant (by 4$\sigma$ on the declination axis) for the components of this system. While we could 
simply dismiss this one candidate out of hand and accept the others as bona-fide binary systems, we have attempted to assess the physical reality of all 17 pairs more quantitatively within a ``Naive''
Bayesian scheme \citep[e.g.][]{zheng98}.  
 
In this approach we adopted the brightest object (at $r$) in each pair as a putative primary star and estimated a Bayes factor for two competing models that can account for the 
secondary component. We assumed it could be either a physical companion $(C)$ or a field white dwarf $(F)$ such that the sum of the probabilities, $P(C)$, it is a companion and $P(F)$,
it is a field star, $P(C) + P(F) = 1$. Considering each candidate binary in turn, we first defined a probability distribution function for the distance of the secondary star, in the case it is a field white dwarf, $P(dist \mid F)$. We divided the visibility cone in the direction of the pair into a series of thin slices (out to 1250pc) and integrated the luminosity function of \cite{degennaro08} over the bolometric magnitude range that is consistent with our survey limits and the mean distance of each slice (adopting log g=8.0
white dwarfs and using the synthetic photometry of the Montreal group to translate between $M_{bol}$ and $M_{r}$) to obtain estimates of the number of white dwarfs per cubic parsec. After factoring in the 300pc 
Galactic scale height of the white dwarf population \citep[e.g.][]{vennes02}, we multiplied the resulting values by the slice volumes to obtain the number of objects within each region along the visibility cone. This distribution function was normalised over the range 0-1250pc, outwith of which it is effectively zero (due to the magnitude limit of our survey).

Next we defined a probability distribution function for the proper motion of the secondary star in the case it is a field white dwarf, $P(pm \mid F)$. We obtained absolute proper motions from the SuperCOSMOS Sky Survey database \citep{hambly01c} for all white dwarfs listed in the SDSS DR7 spectroscopic catalogue \citep{kleinmann13}, meeting our photometric selection criteria and lying within 20$^{\circ}$ of a candidate binary. We modelled the distribution of these proper motions with an asymmetric pseudo-Voigt profile\footnote{www.xray.cz/xrdmlread/PseudoVoigtAsym-m}, normalising this function over the range -150 to +150 mas~yr$^{-1}$ on each axis (the distributions are effectively zero beyond these limits). For the probability distributions of the distance and proper motion of the secondary star, in the case it is a companion white dwarf, $P(dist \mid C)$ and $P(pm \mid C)$, respectively, we adopted normalised Gaussian functions centered on the values observed for the primary component and with widths corresponding to the uncertainties in these measurements. Similarly, for the conditional probabilities of the observed secondary star distances and proper motions, $P(data \mid dist)$ and $P(data \mid pm)$, we adopted normalised Gaussian functions centered on the observed values for the secondary component and with widths matching the uncertainties in those measurements.

\begin{equation}
\label{dataC}
\begin{split}
P(data \mid C) & = \int_{0}^{1250} P(data \mid dist)P(dist \mid C) {\rm d}dist \\
& \int_{-150}^{150}P(data \mid pm)P(pm \mid C)  {\rm d}pm 
\end{split}
\end{equation}

\begin{equation}
\label{dataF}
\begin{split}
P(data \mid F) & = \int_{0}^{1250} P(data \mid dist)P(dist \mid F) {\rm d}dist \\
& \int_{-150}^{150}P(data \mid pm)P(pm \mid F)  {\rm d}pm 
\end{split}
\end{equation}

Subsequently, we used equations~\ref{dataC} and ~\ref{dataF} to calculate $P(data \mid C)$ and $P(data \mid F)$, respectively \citep[e.g.][]{kass95}, for each candidate binary, where
the integrals were evaluated by sampling the functions at several thousand points across their respective ranges. The ratios of these parameters, as listed in the final column of 
Table~\ref{wdmass}, are the approximate Bayes factors of our models. For un-informative priors (ie. $P(C)=P(F)=0.5$) they equate to the ratios of the posterior probabilities of the two models for the secondary, $P(C\mid{\rm data})$ and $P(F\mid{\rm data})$. An inspection of the Bayes factors in Table~\ref{wdmass} reveals that physical association of the components is strongly and very strongly favoured in 2 and 14
pairings, respectively \citep[][]{jeffreys61}. Only the pair we suspected above to be non-physical, CDDS\,34, is rated as 
more likely to be a chance alignment. Consequently, we do not consider it further in this work.

\section{Wide binary white dwarf masses}
\label{massd}
\subsection{The young, wide, double-degenerate mass distribution}

\begin{figure}
\includegraphics[angle=270,width=\linewidth]{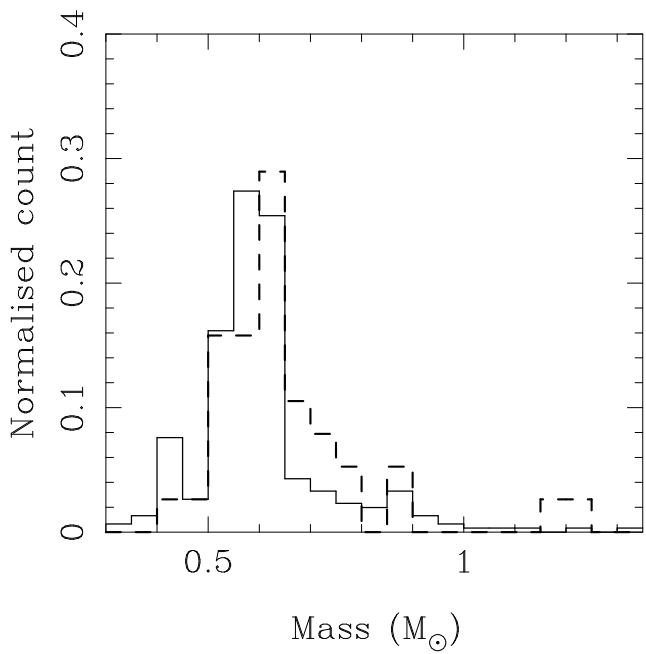}
\caption{Normalised mass distribution for our binary white dwarf sample (heavy dashed line). The normalised mass distribution for the 303 white dwarfs with 
$T_{\rm eff}$$>$12000K in the Palomar-Green Survey is overplotted (thin solid line).}
\label{MD}
\end{figure}

We have expanded our sample of 16 binaries by folding in three previously known hot, wide, non-magnetic, double-degenerates that lie within 
the footprint of SDSS DR7. We have confirmed these to be physical systems by subjecting their components to the same analysis we performed 
on our new candidate binaries (see Table~\ref{wdmass_lit}). Subsequently, we have built a mass distribution from the 38 white dwarfs of 
this enlarged sample (Figure~\ref{MD}). This distribution is of importance because it conveys information relating to the initial mass function
of the progenitor stellar population and the stellar initial-mass final-mass relation \citep[e.g.][]{ferrario05}. 

While the white dwarf mass distribution has been determined several times previously, these efforts have been based on samples dominated by 
isolated field objects e.g. from the SDSS DR4 \citep{kepler07} and from the Palomar-Green (PG) \citep{liebert05a} surveys. These distributions
are loosely described in terms of three peaks, a main one located around $M$=0.6$M_{\odot}$ which arises from the progeny of the numerous F dwarfs 
of the Galactic disk, a high mass one centered at $M$=0.8-0.9$M_{\odot}$, often attributed to binary mergers \citep[e.g.][]{yuan92} but which more
recent kinematic work links to early-type stars that naturally lead to larger remnants \citep{wegg12}, and a low mass one stationed around 
$M$=0.4$M_{\odot}$, associated with close binary evolution and the non-conservative transfer of mass \citep[e.g.][]{marsh95}. It is important to note
that in magnitude limited samples such as those from the PG and the SDSS DR4 surveys, low and high mass white dwarfs are over and under represented 
due to their greater and lesser than average intrinsic luminosities, respectively  \citep[e.g. see][ for a discussion]{liebert05a}. 

It is apparent from Figure~\ref{MD} that the mass distribution of our largely magnitude limited sample of young, wide double-degenerates, also 
displays a strong peak around 0.6$M_{\odot}$. Indeed, we find the mean masses of the 300 PG white dwarfs with $T_{\rm eff}$$>$12500K (we exclude PG0922+162 
and do not attempt to apply our mass corrections at lower temperatures since \citeauthor{liebert05a} used a previous generation of synthetic spectra) and 
ours, 0.60$M_{\odot}$ and 0.65$M_{\odot}$ respectively, to be comparable. Some of the small difference in these means is likely attributable to our use of 
more modern synthetic H-line profiles \citep[see e.g.][]{tremblay11} although unlike the mass distribution of the PG white dwarfs, where there is a 
prominent peak around 0.4$M_{\odot}$ and 37 objects with $M$$<$0.5$M_{\odot}$, our binary sample harbours only two objects with masses that are likely to be
$M$$<$0.5$M_{\odot}$. From drawing 10000 random subsamples of 38 white dwarfs from the PG sample, we estimate the likelihood of there being two or 
fewer objects with $M$$<$0.5$M_{\odot}$ in our sample  as $P$$\sim$0.2, suggesting that, for now, the observed shortfall is not significant. A relative 
deficit of low
mass objects might become apparent in a larger wide, double-degenerate sample because it is likely that low mass products of close binary evolution located
in triple degenerate systems will generally be partnered in wider orbits by white dwarfs of canonical mass. These have smaller radii and are intrinsically 
fainter, reducing the influence of the luminosity bias that operates in favour of the detection of low mass white dwarfs in isolated samples.

Following the same Monte-Carlo approach applied above, we find the presence of two white dwarfs amongst our 19 binaries (CDDS\,18-B and CDDS\,26-A) which sit well above 
recent theoretical predictions for the minimum mass of ONe core degenerates \citep[e.g. $M$$\sim$1.05-1.1M$_{\odot}$, ][]{siess07, gilpons03} to represent a 
significant excess ($P$$\sim$0.05). In the PG sample only 3 out of 300 objects have masses this large. The formation of ultra-massive white dwarfs remains a matter
of some debate since it is unclear if the putative single 
super-AGB progenitor stars \citep[$M_{\rm init}$$>$6M$_{\odot}$ e.g.][]{eldridge04, bertelli09} lose their envelopes sufficiently rapidly to prevent their degenerate 
cores growing until they detonate as electron capture supernovae. Although the ultra-massive field degenerate GD\,50 has been linked with the recent Pleiades star 
formation event \citep[][]{dobbie06b}, it is intriguing that no white dwarfs with masses as large as these two objects have yet been unearthed within open clusters
\citep[][]{dobbie12b, williams09}. Considering their masses are distinctly similar to the sum of the masses of two objects from the main peak in the mass distribution,
alternatively, these white dwarfs might have formed through double-degenerate merging \citep[e.g.][]{vennes97}. While this putative merging process would likely lead 
to discord between the cooling times of the components of these two systems \citep[e.g. see the discussion relating to RE\,J0317-853][]{kulebi10,ferrario97,burleigh99,barstow95}, 
the fundamental parameters of PG\,0922+162A and PG\,0922+162B (CDDS\,18) appear to fit entirely within the framework of standard stellar evolution (see Section~\ref{cool}). 
In the case of CDDS26 coherence between the component cooling times is less satisfactory but there is no overwhelming disparity which flags a more exotic evolutionary 
history for CDDS26-A. 

More broadly, it is of interest that the majority of the higher mass stars in our binary sample are paired with each other. The simplest explanation for this favours 
these objects being the progeny of early-type stars. The white dwarfs within the peak of the mass distribution are typically the end points of stars which have reasonably
long lives on the main sequence (at least several Gyr). The bulk of companion (higher mass) white dwarfs originating from early-type stars with relatively short main 
sequence lifetimes will have cooled below $T_{\rm eff}$$\approx$8000-9000K (e.g. 2.5Gyr for 0.9$M_{\odot}$) and moved outwith our colour selection region prior to the formation
of a double-degenerate system. It is widely documented that stars of greater mass are more likely to reside in multiple systems than those of lower mass. The binary fraction
for the lowest mass M dwarfs is estimated to be only 25-35\% \citep[e.g.][]{delfosse04,reid97}, while it is found to be around 70\% for the O and B stars \citep[e.g.][]{sana08,mason98}. 
However, as the binary fraction of the F and G type stars (the progeny of which dominate the white dwarf mass distribution), is only slightly lower than this at 50-60\% 
\citep[][]{duquennoy91,abt76},
it seems unlikely that gross variations in the binary fraction with stellar mass can account for the excess of very high mass degenerates in our study. Alternatively, it
could have arisen because the projected separation distribution of
companions to early-type stars like the progenitors of these white dwarfs is peaked at several hundred AU compared to a few tens of AU, as for stars of later spectral type 
\citep[][]{derosa13,patience02}. During the final stages of stellar evolution, when the orbits widened, following Jean's theorem \citep[e.g.][]{iben00}, those of the intermediate
mass stars expanded by around a factor 5-10 while those of the F/G stars grew sigificantly less, leaving the peak of the latter systems at or below the resolution limit of
the SDSS imaging data. This may have been further compounded by the bulk of the near equal mass progenitor systems, where both components can evolve beyond the main sequence 
within a Hubble time, populating the lower half of the projected separation distribution \citep[][]{derosa13}.

\subsection{Applying the non-magentic binary sample to probe the origins of unusual white dwarfs: the HFMWDs}

Around 10\% of white dwarfs have a magnetic field with $B$$>$1MG, the HFMWDs. Those with H-rich atmospheres (the majority) are identified from Zeeman splitting of 
the pressure broadened Balmer lines in their optical spectrum. However, the standard spectral technique applied to determine non-magnetic white dwarf masses 
\citep[e.g.][]{bergeron92} does not work well for HFMWDs due to poor understanding of the broadening from the interaction of the electric and magnetic fields at the 
atomic scale. Consequently, mass estimates are available for relatively few HFMWDs and often have large uncertainties. They reveal these objects to be generally 
more massive ($\Delta M \sim40$\%) than their non-magnetic cousins \citep[e.g.][]{wickram05,kepler13}. The difficulties in determining the parameters of HFMWDs, in particular their
ages, also mean that assessing their origins is problematic. 

Short of a detailed statistical analysis of the kinematics of a large sample of hot white dwarfs, age information can only be readily accessed for HFMWDs that
are members of open clusters or specific types of wide binary systems. During the course of this work we have unearthed several new hot, wide, magnetic + non-magnetic 
double-degenerate systems \citep[][Dobbie et al. in prep]{dobbie12a,dobbie13a}. For each system, we have compared our estimate of the mass of the non-magnetic component 
to recent determinations of the stellar initial mass-final mass relation to infer a progenitor mass. This has been matched to a grid of solar metallicity 
stellar evolutionary models to determine the stellar lifetime \citep[][]{girardi00}. Subsequently, we have linearly combined this with the cooling time of the white
dwarf to obtain a limit on the total age of the binary and ultimately a lower limit on the mass of the companion HFMWD's progenitor. 

However, any selection effects imprinted on our young, wide, double-degenerate sample, as discussed above, could potentially lead to biased conclusions 
about the progenitors of HFMWDs, if these systems are examined only in isolation. The two main competing hypothetical pathways to the formation of HFMWDs 
should 
synthesise them on different timescales since stellar lifetimes are strongly mass dependent. For example, if they descend primarily from magnetic
Ap/Bp stars \citep[$M_{\rm init}$$\simgreat$2M$_{\odot}$ e.g.][]{angel81}, then most will be formed within only a few hundred Myr (ie. lifetimes of early-F, A and B stars) after 
the birth of their host population. Alternatively, if their magnetism is generated by differential rotation within gas which envelopes primordial close 
binaries, either when their primary stars ($M_{\rm init}$$\ge$0.8M$_{\odot}$) expand to giant scale and overfill their Roche Lobes 
\citep{tout08, nordhaus11} or during merging of two degenerate remnants \citep{garcia12}, the majority of HFMWDs, like field white dwarfs, will be 
created several Gyr after the birth of their progenitor stars (ie. form mainly from the numerically dominant F and G stars). The mass 
distribution of any wide degenerate companions to HFMWDs, when compared to that of the non-magnetic + non-magnetic white dwarf systems, could provide insight 
on the lifetimes of the progenitor stars and thus clues as to which of these formation pathways is more dominant.

\begin{figure}
\includegraphics[angle=0,width=\linewidth]{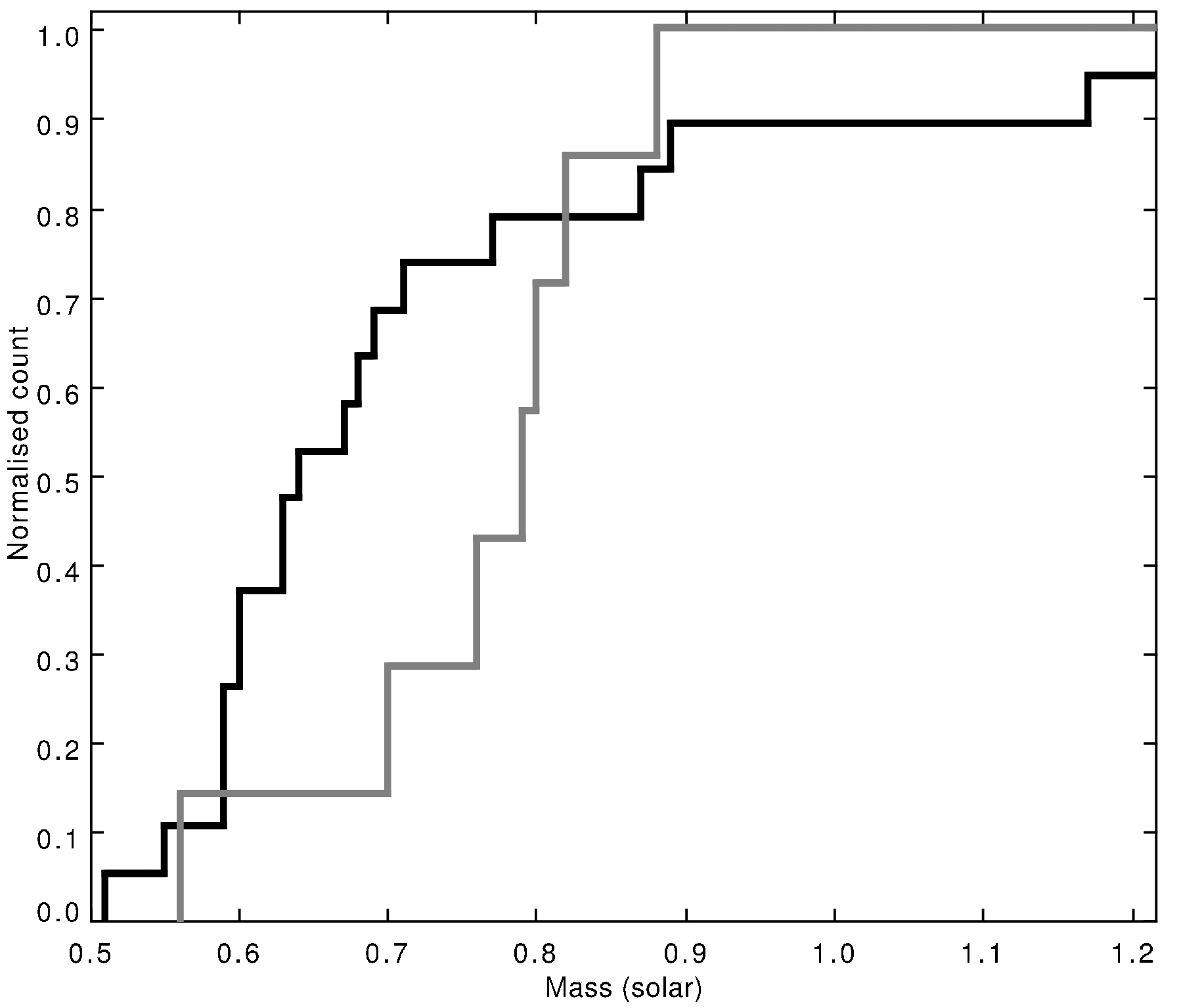}
\caption{Normalised cumulative mass distribution for our binary white dwarf sample (black line). The normalised cumulative mass distribution for the seven DA white dwarfs 
 in magnetic + non-magnetic systems is overplotted (grey line).}
\label{CUM}
\end{figure}

An examination of the component masses in our 19 systems reveals that in 13 cases both are $M$$<$0.7$M_{\odot}$. In contrast, out of seven known hot, wide, magnetic + non-magnetic 
double-degenerates, only one harbours a non-magnetic white dwarf with $M$$<$0.7$M_{\odot}$ \citep[that is the five systems discussed in][and two more reported in Dobbie et al. in 
prep, in which the non-magnetic DA components have masses of 0.76$\pm$0.06$M_{\odot}$, CDDS\,22-A, and 0.56$\pm$0.04$M_{\odot}$, CDDS\,27-A]{dobbie13a}. While the masses of the HFMWDs
in these seven binaries may well have been modified through stellar interactions, the masses of the non-magnetic DA companions are presumably reflective of the spectral types of 
their progenitor stars. Unless some factor has dictated otherwise (e.g. the evolutionary timescales for the formation of the HFMWDs), the progenitor stars of these white dwarfs 
should have had the same broad distribution in mass as those of the components of the 19 non-magnetic + non-magnetic double-degenerate binaries studied here. Adopting this as a null hypothesis, we have drawn thousands of subsets of seven degenerates randomly from the 19 non-magnetic binary systems. We find that only approximately 10 (100) subsamples per 10000 contain six (five) white dwarfs with $M$$\ge$0.7$M_{\odot}$. We have also applied a two-sample Kolmogorov-Smirnov test to the mass distributions of the seven non-magnetic companions and the more massive white dwarf components in each of
our 19 binaries. For $n_{1}$=7 and $n_{2}$=19, the critical $D_{\rm crit}$ values are 0.54 and 0.60 for $P$=0.1 and $P$=0.05, respectively. From the cumulative distributions shown in 
Figure~\ref{CUM} we determine $D$=0.54, which rejects at marginal significance the null hypthesis. These findings suggest these two groups of white dwarfs may have different progenitor mass distributions.

This disparity in these two mass distributions favours a formation model in which the origins of HFMWDs preferentially involve early-type stars, although it does not catagorically rule out 
the hypothesis that they form as a result of stellar interaction. For example, it is possible they are born from higher order stellar systems. Studies of main sequence stars reveal a non-negligible
proportion of binaries are in fact higher order systems, principally triples \citep[$\sim$25\%, ][]{raghavan10}. For reasons of dynamical stability, the majority of these systems are 
structured hierarchically, such that they consist of a relatively close (and potentially interacting) pair with a tertiary companion in a much wider orbit \citep[e.g.][]{black82,tokovinin97}.
Our result would, however, require stars of earlier spectral type to occur more frequently in higher order multiple systems. Moreoever, as current understanding of the IFMR indicates that stars 
of earlier spectral-type naturally lead to more massive white dwarfs, there is less need in this case to invoke merging to account for the larger than average masses of HFMWDs. Admittedly, our
findings appear difficult to reconcile with the almost complete absence of HFMWDs in the more than 2000 close, detached white dwarf + main sequence binaries known to date 
\citep[e.g.][]{rebassa12,parsons13}, particularly considering the existence of a healthy population of magnetic cataclysmic variables. Precision kinematics for large numbers of hot, recently
formed, field white dwarfs from {\it Gaia} \citep{jordan08} will provide a crucial complementary view on the origins of HFMWDs. For now, spectroscopic studies of additional young, wide, double-degenerate systems would be useful to strengthen or refute the trend discussed above.  For example, our initial work could be extended to include objects from the most recent SDSS coverage and regions of sky explored by the SkyMapper \citep{keller07} and VST ATLAS \citep{shanks12} surveys.

\section{The non-magentic binary sample and the IFMR}
\label{cool}

\begin{table*}
\begin{minipage}{160mm}
\begin{center}
\caption{Progenitor masses for the white dwarfs, determined from three recent empirical estimates of the form of the IFMR, 1. \citet{dobbie06a}, 2. \citet{kalirai08} and 3. \citet{williams09}. Stellar lifetimes have been derived from the solar metallicity evolutionary models of \citet{girardi00}. We have calculated the difference between the system age derived from component A and the system age derived from component B. This should be zero for negligible measurement errors and a perfect representation of a monotonic IFMR with no intrinsic scatter. Where an uncertainty is listed as HT, the calculated error bound extends to or lies beyond a Hubble time. When both bounds are larger than a Hubble time we flag the estimate as providing no meaningful constraint (NMC). The projected separation of the components of each system are also shown (rounded to 25AU).}
\label{wdcool}
\begin{tabular}{cccccccc}
\hline
Component & \multicolumn{2}{c}{IFMR 1} & \multicolumn{2}{c}{IFMR 2} &  \multicolumn{2}{c}{IFMR 3} & Separation \\
 & $M$$_{\rm init}$/M$_{\odot}$ & $\Delta$System age/Myr & $M$$_{\rm init}$/M$_{\odot}$ & $\Delta$System age/Myr &  $M$$_{\rm init}$/M$_{\odot}$ & $\Delta$System age/Myr &  /AU \\ \hline

\multicolumn{1}{l}{CDDS3-A} & $2.25^{+0.59}_{-0.52}$ & \multirow{2}{*}{$-901^{+689}_{-2446}$} & $1.78^{+0.46}_{-0.41}$ & \multirow{2}{*}{$-2733^{+3095}_{-HT}$} & $1.93^{+0.34}_{-0.31}$ & \multirow{2}{*}{$-1419^{+1369}_{-5973}$} & \multirow{2}{*}{1725} \\ 
\multicolumn{1}{l}{CDDS3-B} & $1.89^{+0.67}_{-0.60}$ & &  $1.34^{+0.63}_{-0.59}$ &  &  $1.56^{+0.50}_{-0.47}$ & \\ \\

\multicolumn{1}{l}{CDDS6-A} & $2.87^{+0.63}_{-0.55}$ & \multirow{2}{*}{$-155^{+236}_{-276}$} & $2.54^{+0.48}_{-0.46}$ & \multirow{2}{*}{$-216^{+463}_{-546}$} & $2.57^{+0.35}_{-0.34}$ & \multirow{2}{*}{$-188^{+368}_{-412}$}& \multirow{2}{*}{400}\\
\multicolumn{1}{l}{CDDS6-B} & $2.74^{+0.63}_{-0.54}$ & &  $2.39^{+0.48}_{-0.45}$ &  &  $2.44^{+0.36}_{-0.34}$ &\\ \\


\multicolumn{1}{l}{CDDS8-A} & 2.62$^{+0.61}_{-0.52}$ & \multirow{2}{*}{$-630^{+291}_{-515}$} & $2.23^{+0.45}_{-0.42}$ & \multirow{2}{*}{$-736^{+711}_{-1330}$} & $2.31^{+0.33}_{-0.30}$ & \multirow{2}{*}{$-682^{+513}_{-755}$}& \multirow{2}{*}{3225}\\
\multicolumn{1}{l}{CDDS8-B} & 2.53$^{+0.71}_{-0.62}$ & &  $2.13^{+0.63}_{-0.60}$ &  &  $2.22^{+0.50}_{-0.48}$ & \\\\


\multicolumn{1}{l}{CDDS9-A} & 1.33$^{+0.79}_{-0.73}$ & \multirow{2}{*}{$3380^{+HT}_{-3157}$} & $<$1.49 & \multirow{2}{*}{NMC} & $0.99^{+0.68}_{-0.65}$ & \multirow{2}{*}{$11605^{+HT}_{-12393}$}& \multirow{2}{*}{5875}\\
\multicolumn{1}{l}{CDDS9-B} & 2.29$^{+0.70}_{-0.62}$ & &  $1.83^{+0.64}_{-0.60}$ &  &  $1.97^{+0.51}_{-0.48}$ & \\\\


\multicolumn{1}{l}{CDDS14-A} & $2.35^{+0.69}_{-0.61}$ & \multirow{2}{*}{$-618^{+769}_{-1478}$} & $1.91^{+0.62}_{-0.58}$ & \multirow{2}{*}{$-1250^{+2485}_{-11131}$} & $2.04^{+0.49}_{-0.47}$ & \multirow{2}{*}{$-671^{+1143}_{-3637}$}& \multirow{2}{*}{1075}\\
\multicolumn{1}{l}{CDDS14-B} & $2.06^{+0.70}_{-0.64}$ & &  $1.55^{+0.66}_{-0.63}$ &  &  $1.74^{+0.53}_{-0.51}$ & \\\\

\multicolumn{1}{l}{CDDS15-A} &  $1.67^{+0.64}_{-0.59}$ & \multirow{2}{*}{$245^{+3872}_{-2081}$} & $1.08^{+0.59}_{-0.58}$ & \multirow{2}{*}{NMC} & $1.34^{+0.47}_{-0.47}$ & \multirow{2}{*}{$180^{+HT}_{-5726}$}& \multirow{2}{*}{725} \\
\multicolumn{1}{l}{CDDS15-B} &  $1.66^{+0.53}_{-0.48}$ & &  $1.06^{+0.42}_{-0.39}$ &  &  $1.33^{+0.31}_{-0.30}$ &  \\\\

\multicolumn{1}{l}{CDDS16-A} & $<$2.86 & \multirow{2}{*}{NMC} & $<$2.45 & \multirow{2}{*}{NMC} & $<$2.52 & \multirow{2}{*}{NMC}& \multirow{2}{*}{1850} \\
\multicolumn{1}{l}{CDDS16-B} & $1.60^{+0.85}_{-0.72}$ & &  $0.99^{+0.92}_{-0.78}$ &  &  $1.26^{+0.75}_{-0.65}$ &\\\\

\multicolumn{1}{l}{CDDS17-A} &  $3.19^{+0.92}_{-0.85}$ & \multirow{2}{*}{$-273^{+508}_{-809}$} & $2.93^{+1.02}_{-0.99}$ & \multirow{2}{*}{$-834^{+1043}_{-3731}$} & $2.90^{+0.73}_{-0.74}$ & \multirow{2}{*}{$-639^{+830}_{-1377}$}& \multirow{2}{*}{900} \\
\multicolumn{1}{l}{CDDS17-B} &  $2.30^{+0.71}_{-0.64}$ & &  $1.85^{+0.67}_{-0.65}$ &  &  $1.99^{+0.53}_{-0.52}$ &\\\\

\multicolumn{1}{l}{CDDS18-A} & $3.60^{+0.70}_{-0.60}$ & \multirow{2}{*}{$46^{+172}_{-98}$} & $3.42^{+0.52}_{-0.49}$ & \multirow{2}{*}{$96^{+167}_{-100}$} & $3.32^{+0.37}_{-0.36}$ & \multirow{2}{*}{$114^{+128}_{-87}$}& \multirow{2}{*}{525}\\
\multicolumn{1}{l}{CDDS18-B} &  $7.00^{+1.01}_{-0.84}$ & &  $7.58^{+0.61}_{-0.58}$ &  &  $6.83^{+0.31}_{-0.33}$\\\\

\multicolumn{1}{l}{CDDS26-A} & $6.65^{+0.98}_{-0.81}$ & \multirow{2}{*}{$-1535^{+914}_{-5286}$} & $7.15^{+0.62}_{-0.57}$ & \multirow{2}{*}{$-6256^{+4920}_{-HT}$} & $6.47^{+0.34}_{-0.34}$ & \multirow{2}{*}{$-3163^{+1842}_{-HT}$}& \multirow{2}{*}{725} \\
\multicolumn{1}{l}{CDDS26-B} &  $1.76^{+0.66}_{-0.61}$ & &  $1.18^{+0.62}_{-0.60}$ &  &  $1.42^{+0.49}_{-0.47}$ &\\\\

\multicolumn{1}{l}{CDDS29-A} & $4.34^{+0.78}_{-0.66}$ & \multirow{2}{*}{$-72^{+107}_{-204}$} & $4.33^{+0.57}_{-0.53}$ & \multirow{2}{*}{$-176^{+142}_{-250}$} & $4.08^{+0.40}_{-0.38}$ & \multirow{2}{*}{$-164^{+120}_{-180}$}& \multirow{2}{*}{825}\\
\multicolumn{1}{l}{CDDS29-B} & $3.22^{+0.68}_{-0.58}$ & &  $2.96^{+0.53}_{-0.48}$ &  &  $2.93^{+0.37}_{-0.36}$ &\\\\


\multicolumn{1}{l}{CDDS30-A} & $1.61^{+0.66}_{-0.60}$ & \multirow{2}{*}{$946^{+6381}_{-1472}$} & $1.01^{+0.61}_{-0.59}$ & \multirow{2}{*}{NMC} & $1.28^{+0.48}_{-0.47}$ & \multirow{2}{*}{$5495^{+HT}_{-3009}$}& \multirow{2}{*}{425}\\
\multicolumn{1}{l}{CDDS30-B} & $1.97^{+0.67}_{-0.61}$ & &  $1.44^{+0.62}_{-0.59}$ &  &  $1.64^{+0.49}_{-0.47}$ &\\\\

\multicolumn{1}{l}{CDDS31-A} & $4.52^{+0.78}_{-0.68}$ & \multirow{2}{*}{$192^{+101}_{-179}$} & $4.55^{+0.57}_{-0.54}$ & \multirow{2}{*}{$116^{+129}_{-213}$} & $4.27^{+0.39}_{-0.39}$ & \multirow{2}{*}{$118^{+116}_{-165}$}& \multirow{2}{*}{150}\\
\multicolumn{1}{l}{CDDS31-B} &  $3.37^{+0.68}_{-0.59}$ & &  $3.15^{+0.53}_{-0.48}$ &  &  $3.09^{+0.38}_{-0.36}$ &\\\\


\multicolumn{1}{l}{CDDS36-A} & $2.24^{+0.59}_{-0.51}$ & \multirow{2}{*}{$-313^{+574}_{-663}$} & $1.77^{+0.46}_{-0.42}$ & \multirow{2}{*}{$-775^{+1756}_{-3328}$} & $1.93^{+0.34}_{-0.31}$ & \multirow{2}{*}{$-331^{+820}_{-1343}$}& \multirow{2}{*}{4025}\\
\multicolumn{1}{l}{CDDS36-B} & $2.08^{+0.58}_{-0.50}$ & &  $1.57^{+0.45}_{-0.41}$ &  &  $1.76^{+0.33}_{-0.31}$ &\\\\

\multicolumn{1}{l}{CDDS40-A} & $2.33^{+0.68}_{-0.62}$ & \multirow{2}{*}{$138^{+724}_{-393}$} & $1.88^{+0.61}_{-0.60}$ & \multirow{2}{*}{$592^{+2770}_{-869}$} & $2.02^{+0.49}_{-0.48}$ & \multirow{2}{*}{$437^{+1123}_{-695}$}& \multirow{2}{*}{2100}\\
\multicolumn{1}{l}{CDDS40-B} & $2.92^{+0.72}_{-0.65}$ & &  $2.60^{+0.64}_{-0.61}$ &  &  $2.62^{+0.49}_{-0.49}$ &\\\\

\multicolumn{1}{l}{CDDS48-A} & $3.05^{+0.73}_{-0.66}$ & \multirow{2}{*}{$36^{+293}_{-397}$} & $2.75^{+0.64}_{-0.62}$ & \multirow{2}{*}{$-294^{+640}_{-633}$} & $2.75^{+0.49}_{-0.50}$ & \multirow{2}{*}{$-135^{+470}_{-467}$}& \multirow{2}{*}{475}\\
\multicolumn{1}{l}{CDDS48-B} & $2.54^{+0.61}_{-0.53}$ & &  $2.13^{+0.48}_{-0.44}$ &  &  $2.23^{+0.35}_{-0.34}$ &\\\\


\multicolumn{1}{l}{CDDS49-A} & $2.78^{+0.64}_{-0.55}$ & \multirow{2}{*}{$401^{+337}_{-194}$} & $2.43^{+0.50}_{-0.46}$ & \multirow{2}{*}{$623^{+554}_{-325}$} & $2.48^{+0.36}_{-0.35}$ & \multirow{2}{*}{$550^{+381}_{-255}$}& \multirow{2}{*}{525}\\
\multicolumn{1}{l}{CDDS49-B} & $3.58^{+0.70}_{-0.61}$ & &  $3.41^{+0.53}_{-0.50}$ &  &  $3.31^{+0.38}_{-0.38}$ & \\\\

\multicolumn{1}{l}{CDDS50-A} & $2.57^{+0.62}_{-0.54}$ & \multirow{2}{*}{$-347^{+360}_{-663}$} & $2.17^{+0.49}_{-0.45}$ & \multirow{2}{*}{$-551^{+876}_{-2327}$} & $2.26^{+0.36}_{-0.34}$ & \multirow{2}{*}{$-509^{+687}_{-1017}$}& \multirow{2}{*}{375}\\
\multicolumn{1}{l}{CDDS50-B} & $2.37^{+0.70}_{-0.62}$ & &  $1.93^{+0.63}_{-0.59}$ &  &  $2.06^{+0.50}_{-0.48}$ &\\\\

\multicolumn{1}{l}{CDDS51-A} & $1.86^{+0.55}_{-0.48}$ & \multirow{2}{*}{$618^{+1561}_{-594}$} & $1.31^{+0.43}_{-0.41}$ & \multirow{2}{*}{$2800^{+HT}_{-2540}$} & $1.54^{+0.32}_{-0.31}$ & \multirow{2}{*}{$1265^{+2768}_{-1119}$}& \multirow{2}{*}{2200}\\
\multicolumn{1}{l}{CDDS51-B} &  $2.37^{+0.60}_{-0.51}$ & &  $1.93^{+0.47}_{-0.43}$ &  &  $2.06^{+0.35}_{-0.33}$ &\\

\hline

\label{wdcool}
\end{tabular}
\end{center}
\end{minipage}
\end{table*}

Our current knowledge about the form of the fundamentally important stellar initial mass-final mass relation (IFMR) is derived predominantly from 
observations of white dwarfs that are members of open clusters. Most open cluster studies of the IFMR are consistent with a monotonic relation with 
a mild degree of scatter 
\citep[e.g.][]{kalirai08,casewell09,williams09}. Only a modest number of these objects are known to deviate substantially from the general trend 
\citep[e.g. LB\,5893, LB\,390 in nearby Praesepe cluster][]{reid96,casewell09}. However, potentially the open cluster data could have furnished us with
a biased perspective on the form of the relation. While these data now span a comparatively broad range of initial mass ($M_{\rm init}$$\approx$1.5-6$M{
_\odot}$), they are still heavily concentrated between $M_{\rm init}$=2.8-5$M{_\odot}$. Moreover, faint candidate white dwarf members of these populations are
generally identified from their location towards the blue side of colour-magnitude diagrams. Due to the premium on 8m class telescope time any 
photometrically outlying cluster white dwarfs e.g.  cooler, redder objects located closer to the field star population, are arguably less likely to be
targeted for follow-up spectroscopic observations. Studies of white dwarfs in wide binary systems with subgiant and main-sequence stars suggest that there
may be more scatter in the IFMR than indicated by the cluster work \citep{catalan08b,zhao12,liebert13}.

Although our investigation of wide double-degenerate systems cannot provide data points with the absolute age calibration necessary for investigating the 
IFMR in the style of studies of open cluster white dwarfs or white dwarf + subgiant binaries, they can still serve as a useful test of its relative form. 
Hence, we have derived the cooling time for each of our binary white dwarfs, including the DB, with the mixed CO core composition ``thick H-layer'' 
evolutionary calculations of the Montreal Group \citep[][the cooling time of the DB is largely insensitive to our choice of H-layer mass]{fontaine01}. 
The errors on our estimates are derived by propagating the uncertainties in effective temperature and surface gravity.  We have applied three relatively recent, linear model estimates of the IFMR and grids of solar metallicity stellar evolutionary models \citep{girardi00} to infer both the masses and the stellar lifetimes of the white dwarfs' progenitor stars. These IFMRs have been derived by different research teams \citep{dobbie06a,kalirai08,williams09} and do not include white dwarfs with $T_{\rm eff}$$<$12500K, the masses of which can be systematically overestimated \citep[e.g. see][]{kepler07}. Next, we have combined these lifetimes with the degenerate cooling times to obtain two independent 
determinations of the total age of each system for each assumed IFMR. Finally we have calculated the differences between these pairs of age estimates. For 
negligible measurement errors and a perfect model of a monotonic IFMR with no intrinsic scatter, these should be zero. The uncertainties associated with 
these values have been determined following a Monte-Carlo approach.

Examination of the age discrepancies in Table~\ref{wdcool} reveals that importantly, none of the 19 systems is strongly discordant with our current 
understanding of the form of the IFMR. The deviations from zero or no age difference are within 1$\sigma$ for 11 out of the 19 systems and less than about 
2.1$\sigma$ for the remaining binaries, suggesting that the dominant contributors to the non-zero values are the uncertainties on our parameter estimates. 
Unfortunately these are generally large, especially at lower initial masses. This is where the inferred main sequence lifetime is, in an absolute sense, a strong function of mass,
curbing the usefulness of these systems for this work. For initial masses $M_{\rm init}$$\simgreat$2.5$M_{\odot}$ the uncertainties are typically only a few 
100Myr and the parameters of the bulk of systems here can be comfortably reconciled with any of the three model IFMRs. This is perhaps not surprising as
the data around which these models are constructed is drawn primarily from $M_{\rm init}$$\simgreat$2.5$M_{\odot}$. Third dredge-up is also anticipated to be 
rather efficient in this mass regime at preventing further growth of the degenerate core during thermally pulsing asymptotic giant branch evolution 
\citep[e.g. see discussion in ][]{weidemann00}. Any factor that can substantially interfere with the mass loss process at this time, when the radius of a 
star reaches its maximum (e.g. the presence of a close companion) is likely to have a relatively minor impact on the final remnant mass. In comparison, at
lower initial mass, third dredge-up is anticipated to be less efficient, potentially allowing the core to grow significantly during this phase. Differences 
in third dredge-up efficiency, as a function on main sequence mass, might explain some of the additional scatter reported in previous wide binary studies of
the IFMR. These tend to sample lower initial masses. However, some of the extra dispersion in the results of these studies undoubtedly stems from the neglect 
of the spectroscopic overestimate of mass at $T_{\rm eff}$$<$12000K.

\section{Additional notes on some specific systems.} 
\label{tandg}

\subsection{CDDS15}

All new white dwarfs presented here have hydrogen dominated atmospheres, except CDDS15-A which is a DB. Following L151-81A/B \citep{oswalt88}, 
CDDS15 is only the second wide DA + DB system identified to date. However, a number of unresolved, presumably close, DA + DB systems are also 
known or suspected, including MCT\,0128-3846, MCT\,0453-2933 \citep{wesemael94}, PG\,1115+166 \citep{bergeron02,maxted02} and KUV\,02196+2816
\citep{limoges09}. The latter two systems have total system masses close to the Chandrasekar limit (1.4$M_{\odot}$) but although PG\,1115+166 is 
also known to be a post common-envelope binary its components are too widely separated (a$\sim$0.2AU) for it to merge within a Hubble time. 
CDDS15-A and CDDS15-B are both located on the lower side of the peak in the field white dwarf mass distribution and have relatively large tangential
velocities ($v_{\rm tan}$$\approx$90kms$^{-1}$). \cite{dantona91} have suggested that DBs preferentially form for lower mass progenitors ($M_{\rm init}
$$<$2M$_{\odot}$) where the very high mass loss rates experienced at the peak of the last thermal pulse cycle can remove the entire hydrogen surface
envelope in a very short time. However, at least one helium rich white dwarf is known which must, because of its membership of the 625Myr Hyades cluster, 
have descended from a star with $M_{\rm init}$$>$2$M_{\odot}$. L151-81A also appears likely to be the progeny of a moderately early-type star since from 
its higher than average mass ($M$=0.71$M_{\odot}$) hydrogen rich degenerate companion, L151-81B \citep[$T_{\rm eff}$=10460K, log g=8.43][]{gianninas11}, 
we infer an upper limit of $\tau$$\simless$2Gyr on the age of this system. This corresponds to the lifetime of a $M_{\rm init}$$\simgreat$1.7$M_{\odot}$ 
star. Interestingly, spectroscopic surveys of large samples of field white dwarfs have also found the mean mass of the DBs to be significantly larger 
than that of the DA population \citep[e.g.  $\bar{M}_{\rm DB}$=0.67$M_{\odot}$ and $\bar{M}_{\rm DA}$=0.63$M_{\odot}$][]{bergeron11, gianninas11}. 
In a spectroscopic analysis of 140 DBs ($T_{\rm eff}>16000$K) drawn from the SDSS DR7, \cite{kleinmann13} found only one with $M<$0.55$M_{\odot}$. Similarly, 
in a study of 108 relatively bright DBs, 
\cite{bergeron11} also found only one with a mass this low. It seems that CDDS15-A is perhaps slightly unusual for its rather low mass. The construction 
and the investigation of the properties (e.g. separation and companion mass distributions) of a sample of wide DA + DB binaries would likely provide 
additional insight on the evolutionary pathway that leads to the formation of hydrogen deficient degenerates.

\subsection{CDDS30}

The components of CDDS30 are separated by approximately 2.8 arcsec, which translates to around 400 AU at the distance we calculate for this system. It is
unlikely that these two objects have ever exchanged mass via Roche Lobe overflow, so their masses, which are consistent with the lower side of the peak in 
the field white dwarf mass distribution, suggest that they are the progeny of main sequence stars that had comparatively modest masses ($M$$<$2$M_{\odot}$).
We note that this pairing is only 16 arcsec (2400AU at 150pc) from a relatively bright ($B$=13.17, $V$=12.12, $J$=10.37$\pm$0.02, $H$=9.94$\pm$0.03, 
$K_{\rm s}$=9.82$\pm$0.02), late-type star, TYC 2535-524-1, which the Tycho-2 catalogue \citep{hog00} reveals has a similar proper motion ($\mu_{\alpha}\cos
\delta$ = -44.3$\pm$3.5 mas~yr$^{-1}$, $\mu_{\delta}$=30.6$\pm$3.5 mas~yr$^{-1}$). Motivated by this, we have performed a crude estimate of the distance to this 
star. The optical colours drawn from the American Association of Variable Star Observers Photometric All Sky Survey (http://www.aavso.org/apass) appear to 
be consistent with a spectral type of K3--K4 (e.g. HD\,219134, TW Piscis Austri). The near-IR colours from 2MASS are indicative of spectral types K2--K3 
(e.g. $\epsilon$-Eridani, HD\,219134). We are led to conclude that TYC 2535-524-1 is approximately K2--4V, which according to a recently published relation 
between absolute $H$ band magnitude and spectral type \citep{kirkpatrick12}, corresponds to $M_{H}$$\approx$4.0-4.6 mag. From the observed $H$ magnitude of 
$H$=9.94$\pm$0.03 we derive a distance modulus of 5.3--5.9, which is in accord with that we infer for CDDS30-A and CDDS30-B. Considering the similarities
between these distance estimates and the statistically significant proper motion measurements, we are led to conclude that there is a reasonable likelihood 
that these two white dwarfs and TYC 2535-524-1 form a triple stellar system. The mass of the main-sequence star sets a lower limit on the mass of the white 
dwarfs' progenitors of $M$$>$0.6$M_{\odot}$.  However, a detailed spectroscopic study of the early-K dwarf, including activity indicators, may provide limits 
on the total age of this system \citep[e.g.][]{mamajek08} that can lead to better constaints on the progenitor masses.

\subsection{CDDS\,8-A and other possible ZZ Ceti stars}
\label{zzceti}

At least one system in our survey harbours a white dwarf that has been confirmed previously as a pulsating ZZ Ceti star. CDDS\,8-A (SDSS J033236.61-004918.4)
was first reported as a DA in the catalogue of spectroscopically identified white dwarfs compiled from the 1st release of the SDSS data \citep{kleinman04}. It
was subsequently shown to be lying within the instability strip and pulsating with an amplitude of 15.1mma on a period of 767.5s \citep{mukadam04}. As referred 
to in Section~\ref{disto} we find a discrepancy of several hundred $K$ between the effective temperature we measure for this object and that reported by 
\cite{kleinmann13}. It is known that systematic uncertainties of this magnitude in spectroscopic estimates of the atmospheric parameters of ZZ Ceti white dwarfs
can arise if the data are not averaged over several pulsation cycles \citep[e.g. see][]{gianninas11}. Our VLT observations of this object spanned only 1200s, 
corresponding to less than two cycles of the presumably most prominent pulsation mode reported by \cite{mukadam04}. The two distinct SDSS spectra of this star 
also hint at variability in the shape and location of the H$\alpha$ line core. This effect is likely due to velocity fields within the photosphere that result 
from the pulsational motion \citep{koester07j}.

Considering CDDS\,8-A lies on the red edge of the instability strip and has an extremely bright, hot, white dwarf companion, SDSS J033236.86-004936.9, 
\citep[][]{noguchi80,wegner87,mccook99}, from which the distance to and kinematics of this system can be stringently constrained, it represents a particularly 
interesting system for further study. Aficionados of pulsating white dwarfs may be interested in a further 10 components within our
spectroscopic subsample that we have flagged here on the basis of their effective temperatures as either possible (CDDS\,3-A, CDDS\,15-B, CDDS\,16-A, CDDS\,17-B, 
CDDS\,34-B, CDDS\,40-A), probable (CDDS\,29-A, CDDS\,48-A, CDDS\,50-A) or highly probable (CDDS\,14-A) ZZ Ceti stars. A significant proportion of these objects 
are brighter than $g$=18 so could be studied in more detail on telescopes of relatively modest aperture.

\section{Summary}

We have presented spectroscopy for the components of 18 candidate young, wide, double-degenerates photometrically identified within the 
footprint of the SDSS DR7. On the basis of our distance estimates and our astrometry we have concluded that 16 candidates probably form 
physical systems. One of these is a wide DA + DB binary, only the second such system identified to date. We have determined the effective 
temperatures, surface gravities, masses and cooling times of the components of our 16 binaries. We have combined the sample with three 
similar systems previously known from the literature to lie within the DR7 footprint to construct a mass distribution for 38 white dwarfs 
in young, wide double-degenerate binaries. A comparison between this and the mass distribution of the isolated field white dwarf population 
reveals them to have broadly similar forms, each with a substantial peak around $M$$\sim$0.6$M_{\odot}$. However, there is a slight excess of the 
most massive white dwarfs in the binary sample which could be related to the primordial separation distribution of the progenitor systems 
and the expansion of binary orbits during the late stages of stellar evolution. We have shown how our sample can be exploited to probe the 
origins of unusual white dwarfs and found at marginal significance that the progenitor systems HFMWDs are preferentially associated with 
early-type stars, at least within these pairings. Finally we have used the 19 young, wide double-degenerate systems to test the stellar IFMR. 
Within the relatively large uncertainties, no system appears to be strongly discordant with our current understanding of the relation. 

\section*{Acknowledgments}
This work is based in part on observations made with the Gran Telescopio Canarias (GTC), operated on the island of La 
Palma in the Spanish Observatorio del Roque de los Muchachos of the Instituto de Astrofisica de Canarias. The INT and the 
WHT are 
operated by the Isaac Newton Group in the Spanish Observatorio del Roque de los Muchachos of the Instituto de Astrofisica
de Canarias. Based in part on observations made with ESO Telescopes at the La Silla Paranal Observatory under programme 
numbers 084.D-1097 and 090.D-0140. Based in part on observations obtained as part of the VISTA Hemisphere 
Survey, ESO Progam, 179.A-2010 (PI: McMahon). This research was made possible through the use of the AAVSO Photometric All-Sky Survey (APASS),
 funded by the Robert Martin Ayers Sciences Fund. This research has made use of the Simbad database, operated at the Centre 
de Donn\'ees Astronomiques de Strasbourg (CDS), and of NASA's Astrophysics Data System Bibliographic Services (ADS).
Funding for the SDSS and SDSS-II was provided by the Alfred P. Sloan Foundation, with Participating Institutions, the 
National Science Foundation, the U.S. Department of Energy, the National Aeronautics and Space Administration, the 
Japanese Monbukagakusho, the Max Planck Society, and the Higher Education Funding Council for England. The SDSS Web Site 
is http://www.sdss.org/. SDSS is managed by the Astrophysical Research Consortium for the Participating Institutions. 
NL acknowledges support from the national program number AYA2010-19136 funded by the Spanish Ministry of Science and 
Innovation. NL is a Ram\'on y Cajal fellow at the IAC (number 08-303-01-02). Balmer/Lyman lines in the DA models were calculated with the 
modified Stark broadening profiles of \cite{tremblay09}, kindly made available by the authors. We thanks the referee for a constructive report which has improved this paper.
%
%
\bibliographystyle{mn2e}
\bibliography{mnemonic,0195}

\bsp

\label{lastpage}

\end{document}